\documentclass[a4paper,11pt]{article}
\usepackage{graphicx,amsfonts,amssymb}
\topmargin by -\headsep

\newcommand{\la}{\langle}
\newcommand{\ra}{\rangle}

\newcommand{\ds}{\displaystyle}
\newcommand{\avh}{\la H_1 \ra}
\newcommand{\pa}{\partial}
\newcommand{\ud}{{\mathrm d}}
\newcommand{\wb}{\omega}
\newcommand{\N}{\mathcal{N}}
\newcommand{\F}{\mathcal{F}}

\newcommand{\ee}{\epsilon}
\newcommand{\ii}{\mathrm{i}} 
\renewcommand{\d}{\mathrm{d}}
\newcommand{\w}{\omega}
\newcommand{\e}{\epsilon}

\newcommand{\chiZ}{\chi_z}
\newcommand{\chiQ}{\chi_q}
\newcommand{\tchiQ}{\tilde{\chi}_q}
\newcommand{\cn}{\mathrm{cn}}

\newcommand{\dn}{\mathrm{dn}}
\newcommand{\ns}{\mathrm{ns}}
\newcommand{\cs}{\mathrm{cs}}
\newcommand{\middlefig}{.45\textwidth}

\begin{document}

\title{Multibreather and vortex breather stability in Klein--Gordon lattices: Equivalence between two different approaches}
\author{J Cuevas$^1$, V Koukouloyannis$^{2, 3}$, PG Kevrekidis$^4$ and JFR Archilla$^5$\\
\small $^1$ Nonlinear Physics Group of the University of Sevilla, Departamento de F\'isica Aplicada I,\\ \small  Escuela Universitaria Polit\'ecnica, C/ Virgen de \'Africa 7, 41011 Sevilla, Spain\\
\small $^2$ Department of Civil Engineering,\\ \small Technological Educational Institute of Serres, 62124 Serres, Greece\\
\small $^3$ Department of Physics, Section of Astrophysics, Astronomy and Mechanics,\\  \small  Aristotle University of Thessaloniki, 54124 Thessaloniki, Greece\\
\small $^4$ Department of Mathematics and Statistics,\\ \small University of
Massachusetts, Amherst MA 01003-4515\\
\small $^5$ Nonlinear Physics Group of the University of Sevilla, Departamento de F\'isica Aplicada I,\\ \small  ETSI Inform\'atica, Avda. Reina Mercedes s/n, 41012 Sevilla, Spain
}

\date{\today}

\maketitle

\begin{abstract}
In this work, we revisit the question of stability of multibreather
configurations, i.e., discrete breathers with multiple excited
sites at the anti-continuum limit of uncoupled oscillators.
We present two methods that yield quantitative predictions about
the Floquet multipliers of the linear stability analysis around
such exponentially localized in space, time-periodic orbits,
based on the Aubry band method and the MacKay effective Hamiltonian
method and prove that their conclusions are equivalent. Subsequently,
we showcase the usefulness of the methods by a series of case
examples including one-dimensional multi-breathers, and two-dimensional
vortex breathers in the case of a lattice of linearly coupled
oscillators with the Morse potential and in that of the discrete
$\phi^4$ model.
\end{abstract}


\section{Introduction}

Over the past two decades, there has been an explosion of interest
towards the study of Intrinsic Localized Modes (ILMs), otherwise
termed discrete breathers \cite{flach1}. This activity has
been, to a considerable extent, fueled by the ever-expanding
applicability of these exponentially localized in space and periodic
in time modes. A partial list of the relevant applications includes
their emergence in halide-bridged transition metal
complexes as e.g. in \cite{swanson}, their potential
role in the formation of denaturation bubbles in the DNA double
strand dynamics summarized e.g. in \cite{peyrard},
their observation in driven micromechanical cantilever arrays as shown
in \cite{sievers}, their investigation in
coupled torsion pendula \cite{lars}, electrical transmission
lines \cite{lars2}, layered
antiferromagnetic samples such as those of a (C$_2$H$_5$NH$_3$)$_2$CuCl$_4$
\cite{lars3}, as well in nonlinear optics \cite{discreteopt} and
possibly in atomic physics of Bose-Einstein condensates
\cite{mplbkon,morsch} and most recently even in granular crystals
\cite{chiara}.

In parallel to the above experimental developments in this  diverse
set of areas, there has been a considerable progress towards the
theoretical understanding of the existence and stability properties
of such localized modes summarized in a number of reviews and
books; see e.g. \cite{flach1,discreteopt,A97,pgkbook}.
Arguably, one of the most important developments in establishing
the fundamental relevance of this area in coupled nonlinear
oscillator chains has been the work of MacKay and Aubry \cite{macaub},
which established the fact that if a single oscillator has a periodic
orbit (and relevant non-resonance conditions are satisfied), then
upon inclusion of a non-vanishing coupling between adjacent such oscillators,
an ILM type waveform will generically persist.

Given the confirmation of persistence of such modes, naturally, the
next question concerns their robustness under the dynamical evolution
of the relevant systems, which is critical towards their experimental
observability. This proved to be a substantially more difficult
question to answer in a quantitative fashion, especially so
for ILMs featuring multiple localized peaks, i.e., multi-site
breathers (since single-site breathers are typically stable
in chains of linearly coupled anharmonic oscillators). Two principal
theories were proposed for addressing the stability of such
periodic orbit, discrete breather states (and identifying their
corresponding Floquet multipliers). Interestingly, these originated
independently from the same pioneers which established (jointly)
the existence of such modes in \cite{macaub}. In particular,
the first theory was pioneered by Aubry in his seminal work
of \cite{A97} and will go under the name Aubry Band (AB) theory,
hereafter. The second one is an effective Hamiltonian method
which was introduced in a series of papers by MacKay and
collaborators \cite{mackay} (and will be termed accordingly
MacKay Effective Hamiltonian method (MEH)). The AB approach
was adapted to the stability of discrete breathers and
multibreathers in the setting of Klein-Gordon lattices
in the work of \cite{ACSA03}; see also \cite{CAR05}.
The MEH approach was applied to the same setting in the
recent work of  \cite{KK09}; see also \cite{arxiv}.

Our aim in the present work is to unify the two methods
by firmly establishing the equivalence of the stability
conclusions of the Aubry band and MacKay effective Hamiltonian
methods. Subsequently, we illustrate the usefulness and
versatility of the methods, we apply them to a range of
physically interesting chains of oscillator model examples,
such as the Morse potential which arises in the study
of DNA bubbles \cite{peyrard}, as well as the $\phi^4$
potential which arises in applications in dusty plasmas
 \cite{koukour}, as well as in field theory, particle
physics and elsewhere; see e.g. the recent discussion of
\cite{kevcuev} and the earlier review \cite{belova} and references therein.
Our presentation is structured as follows. In section 2,
we compare the two approaches and showcase the equivalence of
their conclusions. In section 3, we study multibreathers
and vortices in the case of the Morse potential. In section
4, we present the corresponding results for the $\phi^4$
Klein-Gordon lattice. Finally, in section 5, we summarize
our findings and present our conclusions.

\section{Comparison between the two approaches}
\subsection{Preliminaries - Terminology}
The relevant system under consideration will be a Klein-Gordon chain of oscillators with
nearest-neighbor interaction and Hamiltonian

\begin{equation}
H=H_0+\e H_1=\sum_{i=-\infty}^{\infty}\left[ \frac{1}{2}p_i^2
+V(x_i)\right]
+\frac{\e}{2}\sum_{i=-\infty}^{\infty}\left(x_i-x_{i-1}\right)^2.
\label{eq:kleingordon}
\end{equation}

As indicated previously, we will examine the two approaches (AB
and MEH) for the linear stability of multi-site breathers of this
general class of systems. Both approaches are based on the notion
of the anti-continuum  limit. In this limit ($\e=0$) we consider
$n$ ``central'' oscillators moving in periodic orbits with the
same frequency $\w$ (this will be our ``multibreather'' for $\e
\neq0$ ), while the rest lie at the equilibrium
$(x,\dot{x})=(0,0)$. For $\e\neq0$ some of these configurations,
depending on the phase differences between the oscillators, are
continued in order to provide multibreather solutions. It is
interesting/relevant to note here that while the MEH approach
provides explicit conditions about which configurations can be
continued to finite $\e$ (the critical points of the relevant
effective Hamiltonian), the AB theory provides only stability
information for a given configuration (for which we already know
otherwise that it should exist at finite $\e$).

The linear stability of these solutions is determined by the
corresponding Floquet multipliers. For a stable multibreather we
require that all the multipliers lie on the unit circle. In the
anti-continuum limit these multipliers lie in three bundles. The
two conjugate ones, that correspond to the non-central
oscillators, lie at $e^{\pm i\w T}$, while the third one lie at
$+1$ and consists of $n$ multiplier pairs, corresponding to the
central oscillators. Each pair of $+1$'s correspond to the {\em
phase mode} and {\em growth mode} of each isolated excited
oscillator, meaning that a small change in the initial phase or a
small change in frequency leads to an extremely  close periodic
solution, for the growth mode with slightly larger or smaller
amplitude.

For $\e\neq0$, the non-central corresponding bundles split and
their multipliers move along the unit circle to form the phonon
band, while the multipliers at unity can move along, either the
unit circle (stability), or along the real axis (instability).
However, a pair of   multipliers always continues at $+1$  corresponding to
the phase mode and growth mode of the whole system.
Hence, the stability of the multibreather, at least for small
values of the coupling, is determined by the multipliers of the
central oscillators. For larger values of $\e$, a Hamiltonian-Hopf
bifurcation can occur and destabilize an initially stable
multibreather.

At this point, it is relevant to make a note in passing about the
striking similarities between the discussion above (at and near
the anti-continuum limit) with that of the linear stability of
standing waves in the discrete nonlinear Schr{\"o}dinger (DNLS)
equation. In that case, due to the monochromatic nature of the
solutions and the U$(1)$ invariance of the latter model, it is
possible to directly consider
the  eigenvalues associated with the
standing wave solutions. However, there is a direct analogy with
the spectrum of the excited sites being associated with the
eigenvalues at the origin at the anti-continuum limit and the
continuous spectrum lying at a finite distance from the spectral
plane origin, and how at finite coupling these zero eigenvalue
pairs of the excited oscillators are the ones that may give rise
to instability. In fact, it turns out that even the conditions
under which instability will ensue for multibreathers of the KG
directly parallel the ones for multi-breathers (or multi-site
standing waves) of the DNLS. The latter are analyzed in
considerable detail for $1-$, $2-$ and $3-$ dimensional settings
in \cite{pgkbook}.

Returning to our KG setting, the MEH approach considers the
Floquet multipliers given as $\lambda=\exp(\sigma T)$, with
$T=2\pi/\w$, whereas in the AB approach,
$\lambda=\exp(\ii\theta)$. Then,
\begin{equation}\label{eq:Floquet}
    \sigma=\frac{\ii\theta}{T}=\frac{\ii\theta\w}{2\pi}
\end{equation}

Due to the symplectic character of the Floquet matrix if $\lambda$
is a multiplier, so is $\lambda^{-1}$, and due to its real
character if a $\lambda$ is non real multiplier, so is
$\lambda^*$, where the asterisk denotes the complex conjugate.
Therefore, the corresponding multipliers come in complex
 quadruplets
 $\left(\lambda, \lambda^{-1}, \lambda^{*},
{\lambda^{*}}^{-1}\right)$ if $|\lambda|\neq 1$ and $\lambda$ is
not real, or in duplets $\lambda,\lambda^{-1}$ if $\lambda$ is
real, or $\lambda,\lambda^{*}$ if $|\lambda|=1$ and not real. In
addition, due to the time translation invariance of the system
there is always a pair of eigenvalues at $+1$. This has as a
result that both $\sigma_i$'s and $\theta_i$'s,  come also in
 quadruplets or duplets
 $\left(\sigma, -\sigma\right)$ if $\sigma$ is real,
($\left(\theta^{*}, -\theta^{*}\right)$ if $\theta$ is imaginary)
or
 $\left(\sigma^{*}, -\sigma^{*}\right)$ if $\sigma$ is imaginary
($\left(\theta, -\theta\right)$ if $\theta$ is real). In
principle, the duplets could collapse at a single value
$\lambda= \pm 1$, but there is always a pair of $+1$'s for the systems
under study, as explained above.

\subsection{The MEH approach}

The MEH approach consists of constructing an effective
Hamiltonian, whose critical points are in correspondence with
periodic orbits (in our case multibreathers) of the original
system. This method has been originally proposed in \cite{mackay} and used in the present form in \cite{koukmac}. The effective Hamiltonian can be constructed
as follows.

After considering the central oscillators we apply the
action-angle canonical transformation to them. Note that, in the
anticontinuous limit, the motion of the central oscillators, in
the action-angle variables, is described by
%
$w_i=\w_it+w_{i0},\ J_i=\mathrm{const.}$, for $i=0,\dots,n-1$.
Where $w_i$ is the angle, $w_{i0}$ is the initial phase and $J_i$
the action of the $i$-th central oscillator. For this kind of
systems, the action of an oscillator can be calculated as
\begin{equation}
    J_i=\frac{1}{2\pi}\int_0^T p_i\,\mathrm{d}x_i=\frac{1}{2\pi}\int_0^T [\dot x_i(t)]^2\,\mathrm{d}t.
\end{equation}


Since we are interested in a first order approach, the effective Hamiltonian can be written as $H^{\mathrm{eff}}=H_0+\e\langle H_1 \rangle$, by neglecting terms which do not contribute to the results in this order of approximation. In this formula, $\langle H_1\rangle$ is the average
value of the coupling term of the Hamiltonian, over an angle in the anticontinuous limit, which is equivalent to the average value of $H_1$ over a period
$$\avh=\frac{1}{T}\oint H_1\ud t.$$
This averaging procedure is performed in order to lift the phase degeneracy of the system. For the same reason we introduce a second canonical transformation
\begin{equation}
\begin{array}{llll}
\vartheta=w_0 & &{\cal A}=J_0+\ldots+J_{n-1}\\
\ds\phi_i=w_i-w_{i-1}& &\ds
I_i=\sum_{j=i}^{n-1} J_j& i=1,\ldots,n-1.\\
\end{array}\label{transformation}
\end{equation}
In these variables, the effective Hamiltonian reads
\begin{eqnarray}
H^{\mathrm{eff}}=H_0(I_i)+\e \avh(\phi_i, I_i).
\label{heff}
\end{eqnarray}
Note that, since the calculations are performed in the anticontinuous limit, the contribution of the non-central oscillators has disappeared.

As we seldom know the explicit form of the transformation
$(x, p) \mapsto (w, J)$, we use the fact that since the motion of the central
oscillators for $\e= 0$ is periodic, and possesses the
$t\mapsto -t, x\mapsto x, p\mapsto -p$ symmetry, it can be described by a
cosine Fourier series
$x_i(t)=\sum_{k=0}^{\infty} A_k(J_i) \cos(k w_i).$

Note that at the anti-continuous limit, the orbits differ only in
phase (i.e. $\w_i=\w\ \forall i$), therefore $J_i=J$ and the coefficients $A_k$'s do not
depend on the index $i$.

So, excluding the constant terms, $\avh$ becomes for the KG
problem \cite{KK09}

\begin{eqnarray}
\avh=-\frac{1}{2}\sum_{k=1}^{\infty}\sum_{i=1}^{n-1}A_k^2\cos(k\phi_i)\label{average}
\end{eqnarray}

One of the main features of the MEH approach
is that the critical points of this effective Hamiltonian
correspond to the multibreather solutions of the system. This fact provides the corresponding persistence conditions, as the simple roots of
$\frac{\pa \avh}{\pa \phi_i}=0$. Remarkably, in this setting,
similarly to what is known also for the DNLS \cite{pgkbook},
it can be proved that the only
available multibreather solutions in the
one-dimensional case are
the ones with relative phase among the excited sites of
$0$ or $\pi$.

The second important fact the MEH approach yields is that the linear
stability of these critical points (i.e., the Hessian of the
effective Hamiltonian) determines the stability of the
corresponding multibreather. In particular, the nonzero
characteristic exponents of the central oscillators $\sigma_i$ (see the discussion in the previous subsection) are
given as eigenvalues of the stability matrix
${\bf E}={\bf J}
D^2H^{\mathrm{eff}}$ where ${\bf J}=\left(\begin{array}{cc}\bf O&-\bf I\\\bf I&\bf O\end{array}\right)$ is the matrix of the symplectic structure.
By using the form in (\ref{heff}) for the $H^\mathrm{eff}$ we get:
\begin{eqnarray}
\bf E=\left(\begin{array}{c|c}\bf A&\bf B\\ \hline\bf C&\bf D\end{array}\right)=\left(\begin{array}{c|c}\e \bf A_1&\e\bf B_1\\ \hline\bf C_0+\e\bf C_1&\e\bf D_1\end{array}\right)=\left(\begin{array}{c|c}
-\e\ds\frac{\pa^2\avh}{\pa\phi_i\pa I_j}&-\e\ds\frac{\pa^2\avh}{\pa\phi_i\pa\phi_j}\\[10pt]
\hline\\[-8pt]
\ds\frac{\pa^2H_0}{\pa I_i I_j}+\ds\e\frac{\pa^2\avh}{\pa I_i\pa I_j}&\ds\e\frac{\pa^2\avh}{\pa\phi_j\pa I_i}
\end{array}\right).\label{matrixE}
\end{eqnarray}
Since the only permitted values of the relative phases are $\phi_i=0$, or $\phi_i= \pi$,
the matrix simplifies considerably acquiring the form:
\begin{eqnarray}
\bf E=\left(\begin{array}{cc}\bf O&\bf B\\\bf C&\bf O\end{array}\right)=\left(\begin{array}{cc}\bf O&\e\bf B_1\\\bf C_0+\e \bf C_1&\bf O\end{array}\right).\label{e1}\end{eqnarray}
which, subsequently, if we consider only the dominant eigenvalue
contributions, we get that $\sigma_i^2=\e\chi_{BC}$, where $\chi_{BC}$ are the eigenvalues of the $(n-1 \times n-1)$ matrix ${\bf B_1}\cdot{\bf C_0}$ which reads
\begin{eqnarray}
{\bf B_1}\cdot{\bf C_0}=-\frac{\pa\w}{\pa J}{\bf Z}=-\frac{\pa\w}{\pa J}\left(\begin{array}{ccccc}
2f_1&-f_1&0 & & \\
-f_2&2f_2&-f_2&0 & \\
 &\ddots&\ddots&\ddots & \\
 & 0 &-f_{n-2}&2f_{n-2}&-f_{n-2}\\
 &  &0 & -f_{n-1}&2f_{n-1}
\end{array}\right).\label{z}
\end{eqnarray}
In this expression $\omega=\partial H_0/\partial J$ denotes the
frequency, while
\begin{equation}\label{eq:fi}
f_i\equiv
f(\phi_i)=\frac{1}{2}\sum_{k=1}^{\infty}k^2A_k^2\cos(k\phi_i).
\end{equation}
This leads to the characteristic exponents (i.e., effective eigenvalues)
of the DB in the form:
\begin{equation}\label{eq:sigma}
    \sigma=\pm\sqrt{-\epsilon\frac{\partial \omega}{\partial J}\chiZ},
\end{equation}
with $\chiZ$ being the eigenvalues of the $(n-1 \times n-1)$ matrix

\begin{equation}\label{eq:Z}
    Z_{i,j}=\left\{
    \begin{array}{l}
    Z_{i,i\pm1}=-f_i \\
    Z_{i,i}=2 f_i \\
    0\ \mathrm{otherwise.} \\
    \end{array}
    \right.
\end{equation}

\subsection{The AB approach}

We demonstrate hereby that (\ref{eq:sigma}) can be reobtained
based on the AB approach, by using the exposition of \cite{ACSA03}.
To this end, we recall that the aim of the AB approach is to look for the
displacement that Aubry's bands \cite{A97} experience when the coupling
$\epsilon$ is switched on. What we plan to do below is to calculate the
Floquet eigenvalues assuming that the bands are parabolic and their shape does
not change when the coupling is introduced.

First, we recall the basics of Aubry's band theory with the
notation used in \cite{ACSA03}
adapted to the notation in the present paper, where convenient,
for ease of comparison.
The Hamilton equations applied to the Hamiltonian of
Eq.~(\ref{eq:kleingordon}) can
be written as:

\begin{equation}
\ddot{x}_n \,+\,V'(x_i)\,+\, \ee \,\frac{\partial H_1}{\partial
x_i} \,=\,0\,\quad i=1,\dots,N\,, \label{eq:kleinH1}
\end{equation}
for a generic coupling potential $H_1$, or, if it is harmonic:
\begin{equation}
\ddot{x}_n \,+\,V'(x_i)\,+\,\ee \,\sum_{i=1}^N \,C_{i\,j}x_j
\,=\,0\,\quad i=1,\dots,N \label{eq:klein}
\end{equation}
where $C$ is a coupling constant matrix. Let us define $x\equiv
[x_1(t), \dots,x_N(t)]^\dag$ ($^\dag$ meaning the transpose
matrix). Defining $V(x)=[V(x_1),\dots,V(x_N)]^\dag$, $ {\partial
H_1}/{\partial x }=[{\partial H_1}/{\partial x_1},\dots, {\partial
H_1}/{\partial x_N} ]^\dag$ and so on, Eq.~(\ref{eq:kleinH1}) can be
written as:
\begin{equation}
\ddot{x}  \,+\,V'(x) \,+\,\ee \, \frac{\partial H_1}{\partial x} \
\,=\,0\,.
 \label{eq:kleinket}
\end{equation}
Suppose that $x(t)$ is a time--periodic solution, with period $T$
and frequency $\wb$, its (linear) stability depends on the
characteristic equation for the Newton operator $\N_\ee$ given by

\begin{equation}
 \N_\ee(u)\, \xi   \,\equiv \,\ddot{\xi}  \,
+\,V''(x)* \xi \,+\,\ee\,\frac {\partial^2 H_1}{\partial x^2} \,\xi
=\,E\,\xi \,,
 \label{eq:gnewton}
\end{equation}
where $*$~product is the list product, i.e., $f(x)* \xi $ is the
column matrix with elements $f(x_i(t))\,\xi_i(t)$, and
${\partial^2 H_1}/{\partial x^2}$ is the matrix of functions
${\partial^2 H_1}/{\partial x_i\partial x_j}$, which depends on $t$
through $x=x(t)$.

If $E=0$, this equation describes the evolution of small
perturbations $\xi=\xi(t) $ of $x=x(t) $, which
determines the stability or instability of $x$. It is however,
extremely useful to consider the characteristic equation for any
eigenvalue $E$ as it is the cornerstone for Aubry's band theory.

 Any solution $\xi$  of Eq.~(\ref{eq:gnewton}) is
determined by the column matrix of the initial conditions for
positions and momenta $\Omega(0)=[\xi_1(0),\dots
\xi_N(0),\pi_1(0),\dots \pi_N(0)]^\dag$, with
$\pi_i(t)=\dot{\xi}_i(t)$. A basis of solutions is given by the $2\,N$ functions with initial
conditions $\Omega^\nu(0)$, $\nu=1,\dots,2\,N$, with
$\Omega^\nu_l(0)=\delta_{\nu\,l}$.

The Newton operator depends on the $T$--periodic solution $x(t)$,
and therefore, it is also $T$-periodic and its eigenfunctions can be
chosen also as eigenfunctions of the operator of translation in
time (a period $T$). They are the Bloch functions $\xi(\theta_i,t)=\chi(\theta_i,t) \,
\exp(\ii \,\theta_i \, t/T)$,
 $\chi(\theta,t)$ being a column matrix of $T$--periodic functions. The
sets $\{\xi(\theta_i,0),\dot\xi(\theta_i,0)\}$ are also the eigenvectors
of the Floquet operator $\F_E$ or monodromy, that maps $\Omega(0)$
into $\Omega(T)$, that is, $\Omega(T)=\F_E\Omega(0)$. Their
corresponding eigenvalues are the $2N$ multipliers
$\{\lambda_i\}=\exp(\theta_i)$, with $\{\theta_i\}$ being the $2N$
Floquet arguments.

The set of points $(\theta,E)$, with $\theta$ being a real Floquet
argument of $\F_E$, has a band structure. As the Newton and Floquet
operators are real, the Floquet multipliers come in complex conjugate pairs. Therefore, if $(\theta,E)$ belongs to a
band (i.e. $\theta$ is real), $(-\theta,E)$ does it too, i.e., the
bands are symmetric with respect to $\theta$, which implies that
$\d E/\d \theta(0)=0$. There are always two $T$-periodic solutions,
with Floquet multiplier $\lambda=1$ ($\theta=0$) for $E=0$. One is
$\dot{x}(t)$, which represents a change in phase of the
solution $x(t)$ and it is called the {\em phase mode}; the other
is called the {\em growth mode}, given by $\partial
x(t)/\partial \wb$, and represents a change in frequency and
consequently in amplitude. The consequence is that there is always
a symmetric band tangent to the axis $E=0$ at $\theta=0$.

There are at most $2N$ points for a given value of $E$ and,
therefore, there are at most $2\,N$ bands crossing the horizontal
axes $E=0$ in the space of coordinates $(\theta,E)$. The condition
for linear stability of $x(t)$ is equivalent to the existence of
$2\,N$ bands crossing the axis $E=0$ (including tangent points
with their multiplicity). If a parameter like the coupling $\ee$
changes, the bands evolve continuously, and they can lose
crossing points with $E=0$, leading to an instability of the system.

The first item to find out are the bands at the anticontinuous
limit, where Eq.~(\ref{eq:klein}) reduces to $N$ identical
equations:
\begin{equation}
 \ddot{x}_i \,+\,V'(x_i)\,=\,0\, .
\label{eq:dyn0}
\end{equation}
If we consider solutions around a  minimum of $V$, the oscillators
can be at rest $x_i\,=\,0$, or oscillating with period $T$; the
latter are identical except for a change in the initial phase, so
they can be written as $x_i(t)=g(\w t+w_{i0})$ with $g(\w t)$
being the only $T$-periodic, time-symmetric solution of
Eq.~(\ref{eq:dyn0}) with $g(0)>g(\pi)$. Therefore, the excited
oscillators can be written as:
\begin{equation}
    x_i(t)=z_0+2\sum_{k=1}^\infty z_k\cos[k(\w t+w_{i0})]=\sum_{k=0}^\infty A_k\cos[k(\w
    t+w_{i0})]=\sum_{k=0}^\infty A_k\cos(k w_i) \,,
\label{eq:xizk}
\end{equation}
with $A_k=2z_k$ if $k>0$, $A_0=z_0$ and $w_i=\omega t+w_{i0}$.

Let $n$ be the number of excited oscillators at the anticontinuous
limit, labeled $i=0,\dots,n-1$. Then, there are $n$ identical
bands tangent to the axis $E=0$ at $\theta=0$ for each excited
oscillator, and $N-n$ bands, corresponding to the oscillators at
rest, with $2(N-n)$ points intersecting the $E=0$ axis.

Thus, the excited bands can be approximated around
$(\theta,E)=(0,0)$ by
\begin{equation}
    E(\theta)\approx E_0+\kappa\theta^2 \, ,
    \label{eq:bandshape}
\end{equation}
with $E_0=\epsilon \chiQ$ and $\chiQ$ being the eigenvalues of the $(n\times n)$ $Q$-matrix defined below. Additionally,
\begin{equation}\label{eq:kappa}
    \kappa=\frac{1}{2}\frac{\partial^2 E}{\partial \theta^2}=-\frac{\w^2}{4\pi^2J}\frac{\partial H}{\partial \w}=-\frac{1}{T^2J}\frac{\partial H}{\partial \w}
\end{equation}
where we have made use of \cite[Eq. (B14)]{ACSA03}. The factor
$\kappa$ is positive if the on-site potential $V$ is hard and
negative if $V$ is soft (a potential is hard if the oscillation
amplitude increases with the frequency and soft otherwise). When the
coupling is switched on, the bands will move and change shape; the
$E=0$ eigenvalue is degenerate with multiplicity $N-n$ at $\ee=0$,
but this degeneracy is generically lifted for $\ee\neq 0$ and only
one band will continue being tangent at $(\theta,E)=(0,0)$ due to the
phase mode. Applying degenerate perturbation theory to
Eq. (\ref{eq:gnewton}), with $\ee H_1$ being
the perturbation, a perturbation matrix
$Q$ can be constructed~\cite{ACSA03}, whose eigenvalues $\chiQ$
are those of the perturbed Newton operator. The non-diagonal elements of $Q$ are given by
\begin{equation}
Q_{i\,j}\,=\,\frac{1}{\mu_i\,\mu_j}\int_{0}^{T}\,\dot{x}_i
\frac{\partial^2 H_1 }{\partial x_i\,\partial x_j}\,
\dot{x}_j\,\d\,t\,, \quad i\neq j,\,
\quad i=0\dots n-1,\quad j=0\dots n-1,\,
 \label{eq:qnm}
\end{equation}
 with $\mu_i=\sqrt{\int_{0}^{T}(\dot{x}_i)^2\d t}$.
The diagonal elements are
\begin{equation}
Q_{i\,i}=-\sum_{j\neq i}\,\frac{\mu_j}{\mu_i}\,Q_{i\,j}
\,.
 \label{eq:qnn}
\end{equation}
If the on-site potential $V(x_i)$ is homogeneous and the coupling
is given as in Eq.~(\ref{eq:kleingordon}), as is the case in the
present paper, $\mu_i=(2\pi J)^{1/2}$ $\forall i$. Let us
calculate the derivatives of $H_1$,
 $h_{i,j}= {\partial^2 H_1 }/{\partial x_i\,\partial x_j}$. Because of the
 way the diagonal elements of $Q$ are constructed, we only need the derivatives
 with $i\neq j$. It is easy to see that they are zero except for
 $h_{i-1,i}=h_{i,i-1}=-q_i$ (defined below) for
$i=1,\dots,n-1$. The derivatives $h_{0,n-1}$ and $h_{n-1,0}$ are
also zero as the oscillators at the extremes of the multibreather
are not coupled between them. Then, the matrix $Q$ becomes:
\begin{equation} \label{eq:Q}
    Q_{i,j}=\left\{
    \begin{array}{l}
    Q_{i,i-1}=Q_{i-1,i}=-q_i,\quad \mathrm{for}\, i=1\dots n-1 \\
    Q_{i,i}= q_{i-1}+ q_{i},\quad \mathrm{for}\, i=1\dots n-2 \\
    Q_{0,0}= q_1\\
    Q_{n-1,n-1}= q_{n-1}  \\
    0\ \mathrm{otherwise} \\
    \end{array}
    \right.
\end{equation}
\noindent or, explicitly:

\begin{eqnarray}
Q= \left(\begin{array}{ccccc}
q_1&-q_1&0 & & \\
-q_1&q_1+q_2&-q_2&0 & \\
 &\ddots&\ddots&\ddots & \\
 & 0 &-q_{n-2}&q_{n-2}+q_{n-1}&-q_{n-1}\\
 &  &0 & -q_{n-1}& q_{n-1}
\end{array}\right),\label{matrixQ}
\end{eqnarray}

\noindent with

\begin{equation}\label{eq:qi}
    q_i\equiv q(\phi_i)=\frac{\int_0^T \dot x_i(t)\dot x_{i-1}(t)\,\mathrm{d}t}
    {\int_0^T [\dot x_i(t)]^2\,\mathrm{d}t}=
    \frac{\w}{2 J}\sum_{k\geq1} k^2 A_k^2\cos(k\phi_i)=\frac{\w}{J}f_i,\quad
    i=1,\dots n-1,\,,
\end{equation}

Then, by using \cite[Lemma 5.4]{SAN98} we see that the matrices
$Q$ and $\displaystyle \frac{\w}{J} Z$ have the same nonzero
eigenvalues i.e.
\begin{equation}\label{eq:chi}
    \chiQ=\frac{\omega}{J}\chiZ \,.
\end{equation}
In addition, $Q$ has also a zero eigenvalue.

Some important values of $q(\phi)$ are the following ones:
\begin{eqnarray}
    q(0) &=& 1 \\
    q(\pi) &=& \frac{\sum_{k\geq1}(-1)^k k^2z_k^2}{\sum_{k\geq1}k^2z_k^2}\equiv -\gamma
\end{eqnarray}

For a Morse potential, $\gamma=\w$; for an even potential, $\gamma=1$.

According to the AB theory \cite{A97}, the
Floquet multipliers are given by the cuts of the bands with the $E=0$ axis;
thus

\begin{equation}
    \theta=\pm\sqrt{-\frac{E_0}{\kappa}}=\sqrt{-\frac{\epsilon\w}{\kappa J}\chiZ}
\end{equation}

and, applying the last results
\begin{equation}\label{eq:theta}
    \theta=\pm T\sqrt{\epsilon\frac{\partial \w}{\partial J}\chiZ},
\end{equation}

where we have taken into account that \footnote{The expression $\w=\partial H/\partial J$ comes from the Hamilton's equations for the action-angle variables, since all the calculations are performed in the uncoupled, and therefore integrable, limit.}

\begin{equation}
    \frac{\partial H}{\partial \w}=\frac{\partial H}{\partial J}\frac{\partial J}{\partial\w}=\w\frac{\partial J}{\partial\w}
\end{equation}

Finally, introducing (\ref{eq:theta}) into (\ref{eq:Floquet}), we get (\ref{eq:sigma}),
which completes the proof of equivalence of the relevant Floquet
multiplier predictions.

\section{The Case Example of the  Morse potential}

\subsection{Characteristic exponents}

We now consider some special case examples, starting with
a linearly coupled lattice of oscillators subject to the Morse potential.
As indicated previously, the only configurations that may
exist in the one-dimensional setting are ones which involve
excited oscillators either in-phase (i.e., with $\phi_i=0$)
or out-of-phase (i.e., with $\phi_i=\pi$); see \cite{CAR05},
\cite{KK09} and also \cite{Cuevas} for a detailed discussion.
Here, we
proceed to perform some explicit calculations for the Floquet
multipliers $\sigma$ in the case of n-site breathers. In what follows we
consider only the positive $\sigma$. To this end we express (\ref{eq:sigma})
making use of (\ref{eq:chi}):\footnote{In what follows, and in order
to fix ideas, given the equivalence of the two methods, we will
use the formulation with the $Q$-matrix.}

\begin{equation}
    \sigma=\sqrt{-\epsilon\frac{J}{\w}\frac{\partial \omega}{\partial J}\chiQ(\phi)}
\label{floq}
\end{equation}

where $\chiQ(\phi)$ denotes the Q-matrix eigenvalues for a given $\phi$.
It is straightforward to show
that\footnote{We are neglecting the 0 eigenvalue, associated with
$m=0$.}

\begin{equation}\label{eq:chiQ0}
    \chiQ(0)=4\sin^2\frac{m\pi}{2n}\qquad m=1,\ldots,n-1
\end{equation}

and that

\begin{equation}\label{eq:chiQpi}
    \chiQ(\pi)=-4\gamma\chiQ(0)
\end{equation}

For instance, in the case of a 2-site breather, $\chiQ(0)=2$ and $\chiQ(\pi)=-2\gamma$.

We now focus on the particular case of the Morse potential, since it is
a potential for which closed form analytical expressions can be found.
[For other types of potentials, some approximations can be made for small and
high frequencies; alternatively, the required single-oscillator
parameters, such as
$J$ and $\partial \omega/\partial J$ can be calculated numerically].

In the Morse case and in order to evaluate $J$
and $\partial \omega/\partial J$, we express $J$ as a function of the
Fourier coefficients:

\begin{equation}
    J=2\w\sum_{k\geq1}k^2z_k^2.
\end{equation}

For this  potential,

\begin{equation}
    z_0=\ln\frac{1+\w}{2\w^2};\qquad z_k=\frac{(-1)^k}{k}r^{k/2},\ r=\frac{1-\w}{1+\w}.
\end{equation}

Substituting into the action

\begin{equation}
    J=1-\w \rightarrow \frac{\partial \w}{\partial J}=-1.
\end{equation}

Thus,

\begin{equation}
    \sigma(\phi)=\sqrt{\epsilon\frac{1-\w}{\w}\chiQ(\phi)}.
\end{equation}

In the case of a general phase, we can express $\chiQ(\phi)=q(\phi)\chiQ(0)$ with $q(\phi)$ given by (\ref{eq:qi}). In the special case of the Morse potential,
we have

\begin{equation}\label{eq:qphi}
    q(\phi)=\frac{2\w}{J}\sum_{k\geq1}r^k\cos(k\phi).
\end{equation}

To obtain the relevant sum, we use a simple geometric series formula that can
be found e.g. in \cite{GR65}, according to which:

\begin{equation}\label{eq:qphimorse}
    q(\phi)=\frac{2\w}{J}r\frac{\cos\phi-r}{1-2r\cos\phi+r^2}.
\end{equation}

Consequently,

\begin{equation}
    \sigma(\phi)=\sqrt{2\epsilon r\frac{\cos\phi-r}{1-2r\cos\phi+r^2}\chiQ(0)}.
\end{equation}

For the relevant values of $\phi$ for time-reversible multibreathers, we get:

\begin{equation}
    \sigma(0)=\sqrt{\epsilon\frac{1-\w}{\w}\chiQ(0)}=2\sin\frac{m\pi}{2n}\sqrt{\epsilon\frac{1-\w}{\w}}\qquad m=1,\ldots,n-1
\end{equation}

\begin{equation}
    \sigma(\pi)=\sqrt{-\epsilon(1-\w)\chiQ(0)}=2\sin\frac{m\pi}{2n}\sqrt{-\epsilon(1-\w)}\qquad m=1,\ldots,n-1
\end{equation}

\begin{figure}[t]
\begin{center}
\begin{tabular}{cc}
    \includegraphics[width=\middlefig]{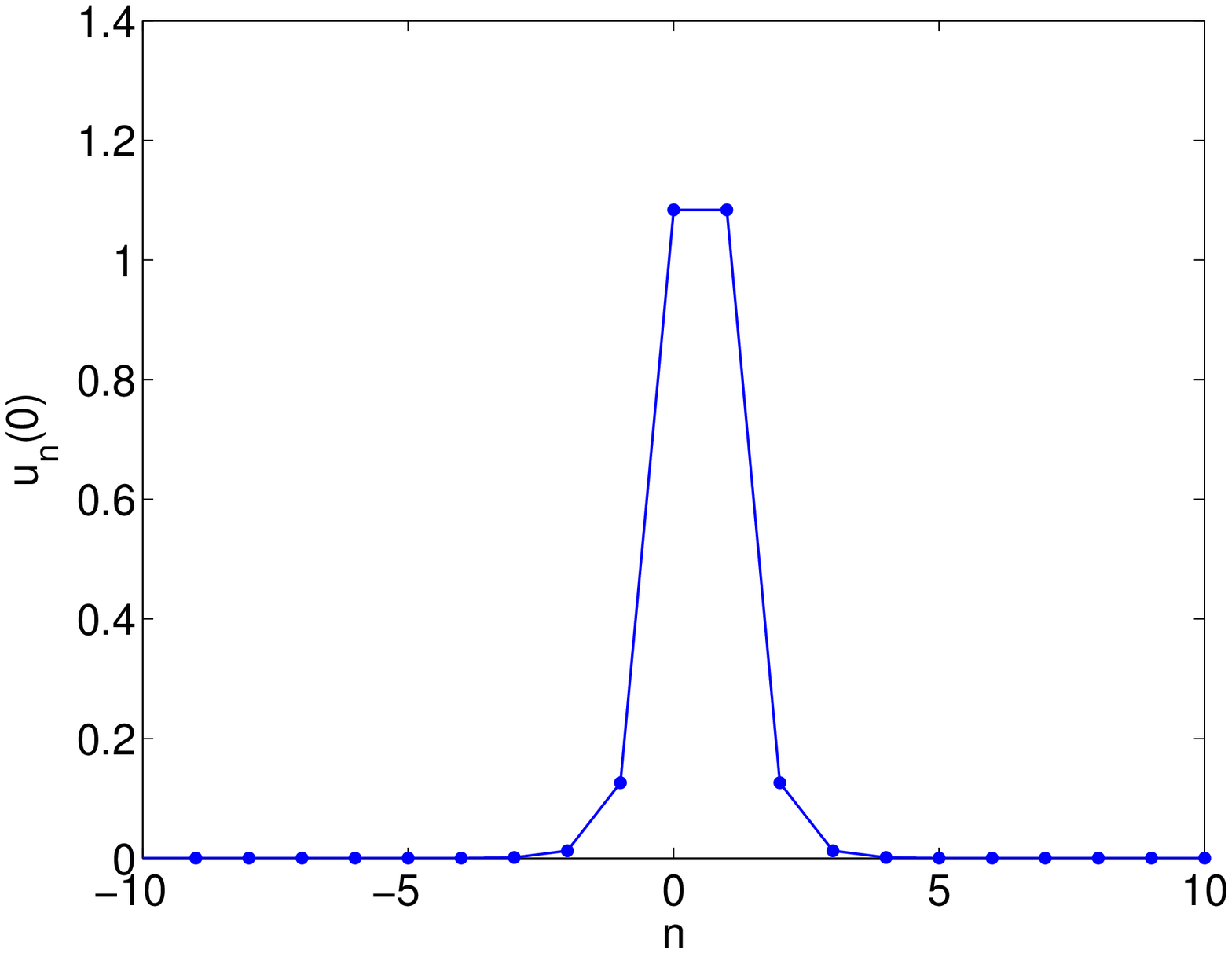} &
    \includegraphics[width=\middlefig]{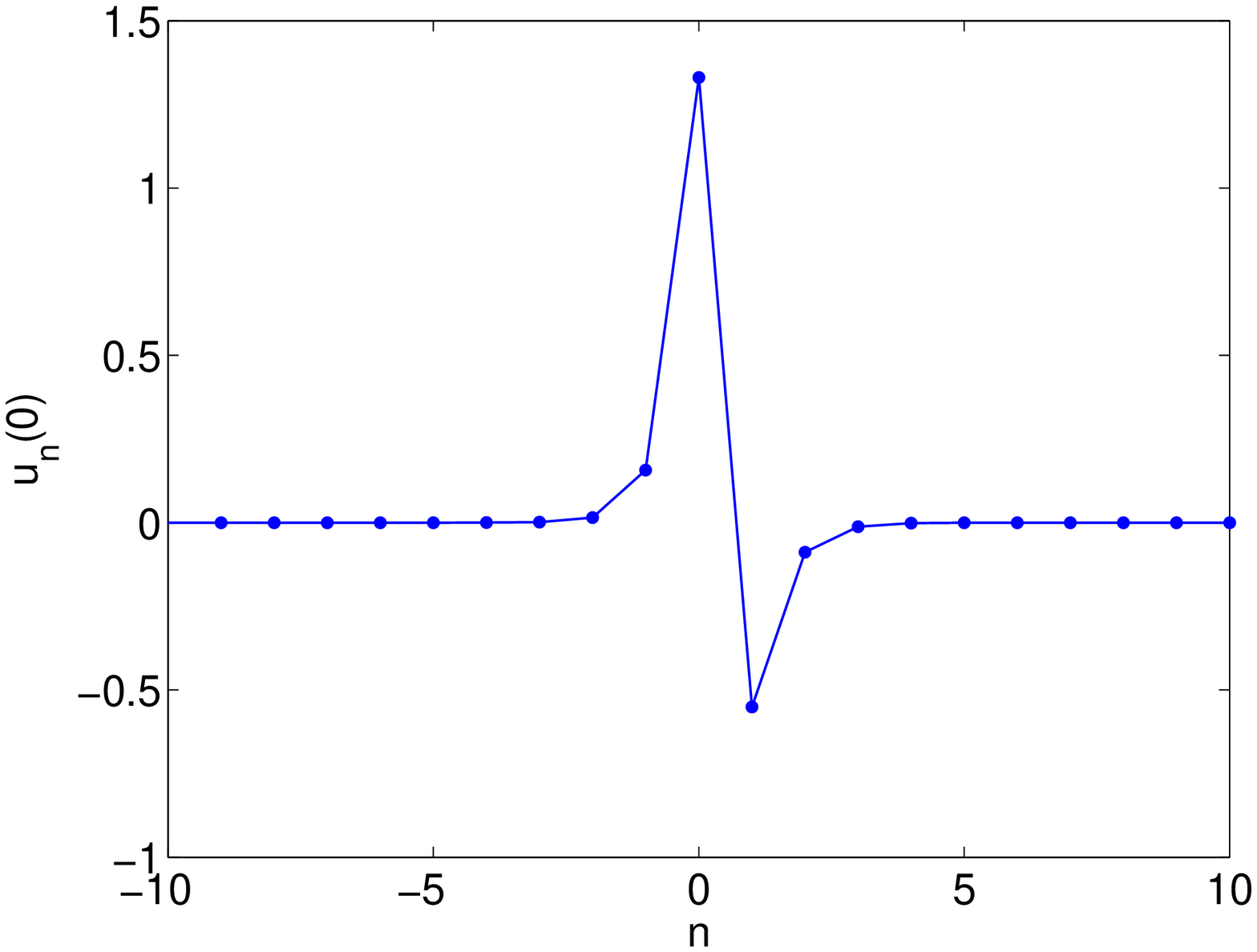} \\
    \includegraphics[width=\middlefig]{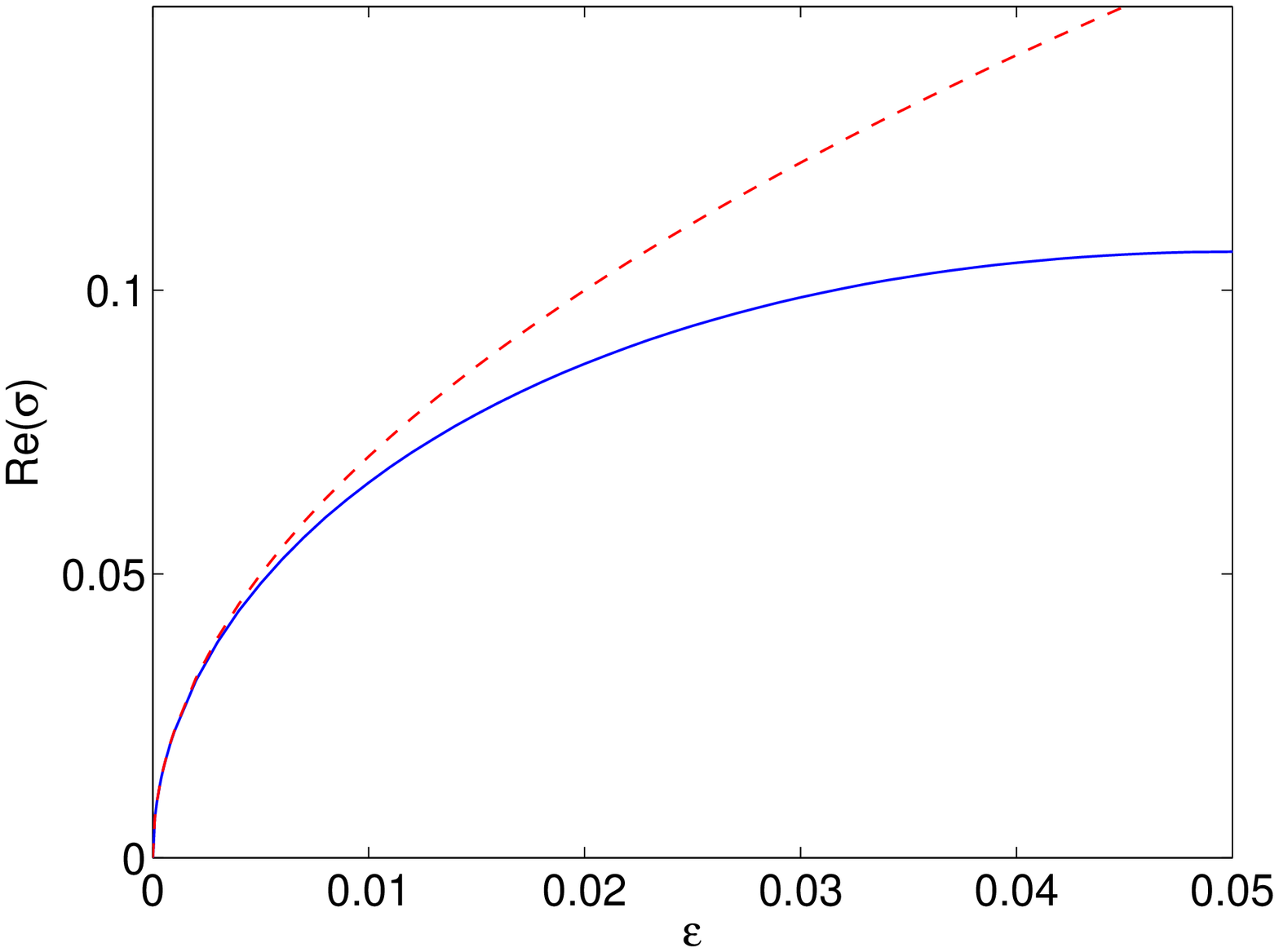} &
    \includegraphics[width=\middlefig]{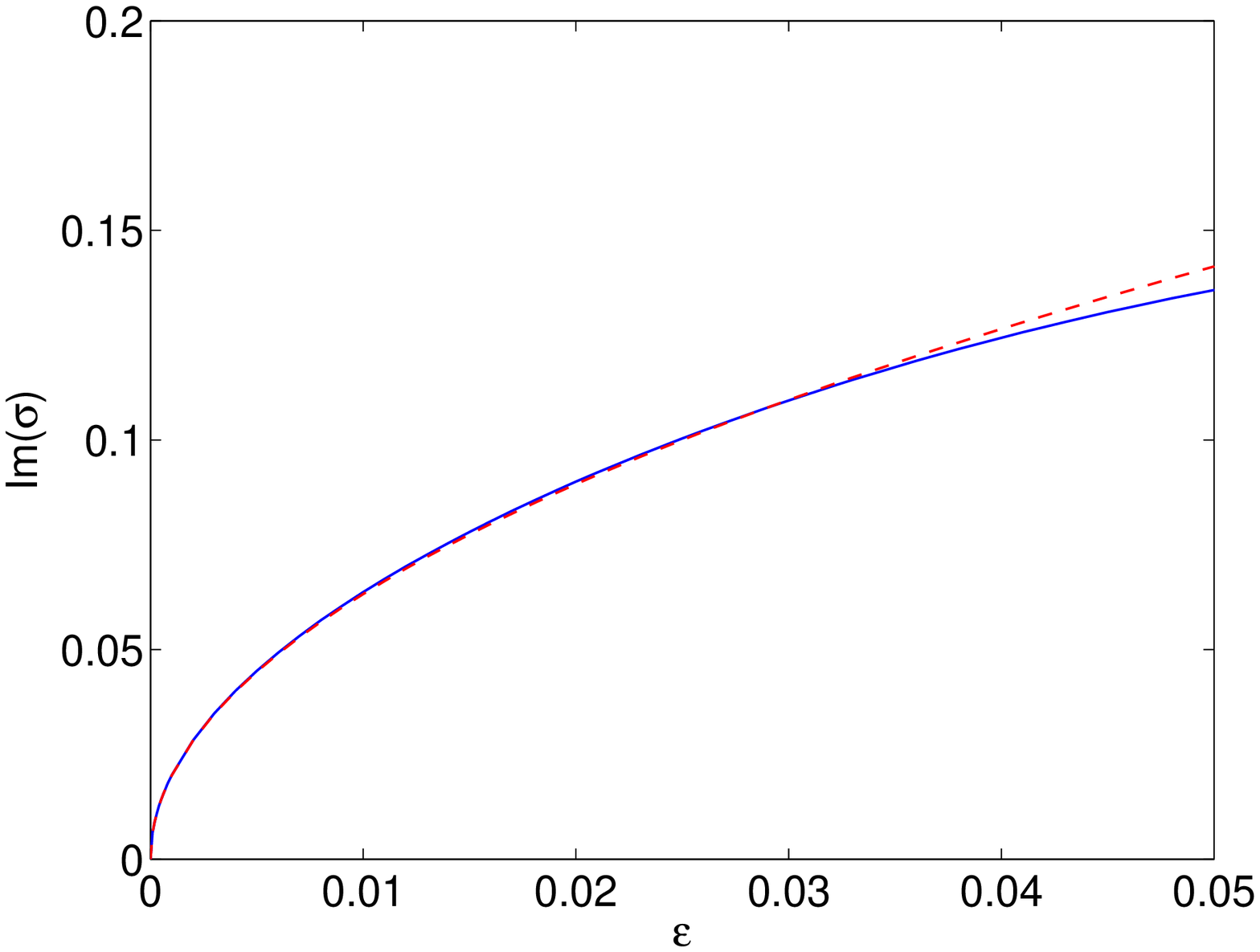} \\
\end{tabular}
\caption{(Top panels) Profiles of an in phase (left) and an out-of-phase
(right) 2-site breather with the Morse potential
for $\w=0.8$ and $\epsilon=0.05$. The bottom panels show the value of the
characteristic exponents $\sigma$ of the corresponding configurations,
with respect to the coupling parameter $\e$. Dashed lines correspond to the
predictions of the stability theorems, while solid ones to full
numerical linear stability analysis results.} \label{fig:2site}
\end{center}
\end{figure}

\begin{figure}[t]
\begin{center}
\begin{tabular}{cc}
    \includegraphics[width=\middlefig]{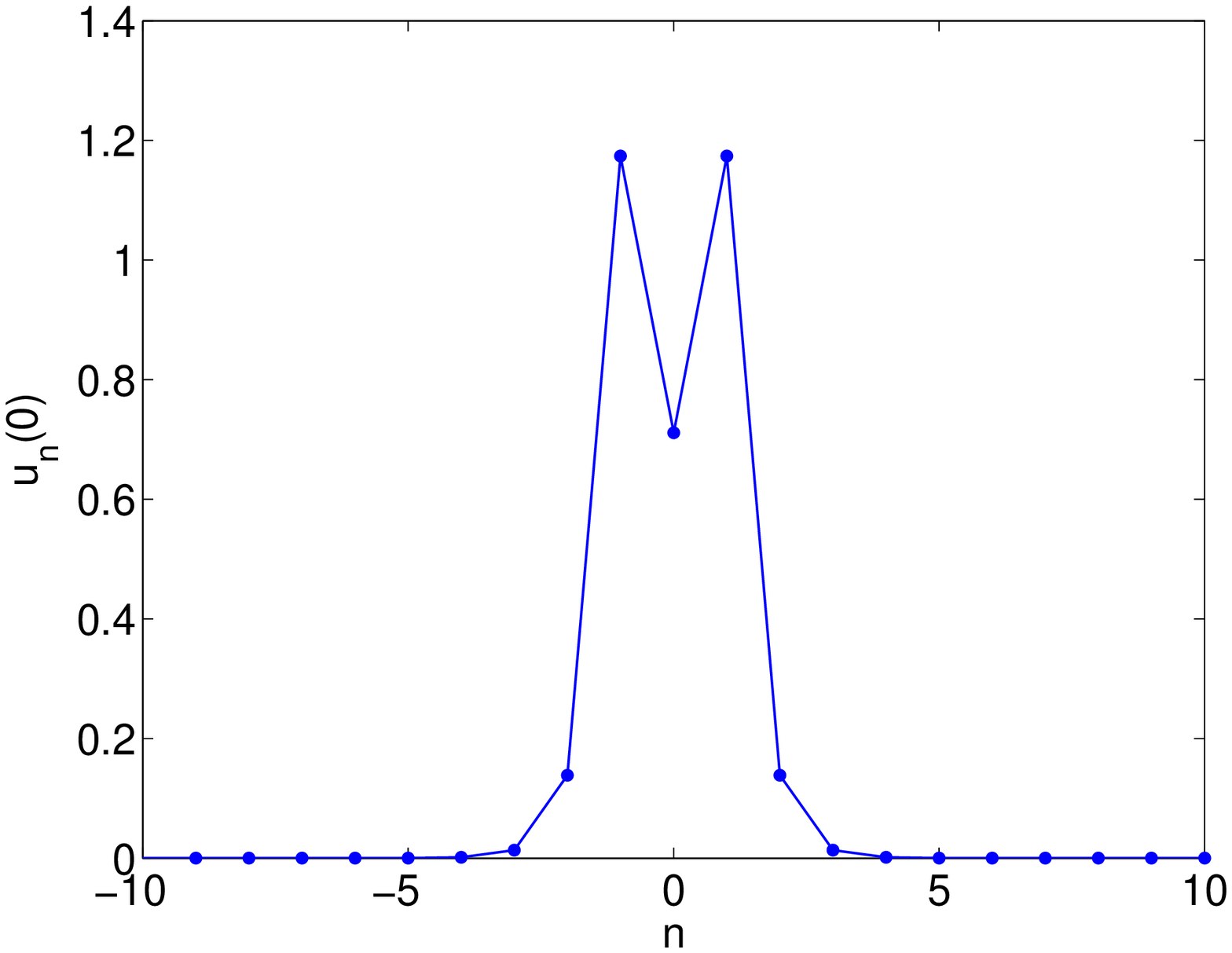} &
    \includegraphics[width=\middlefig]{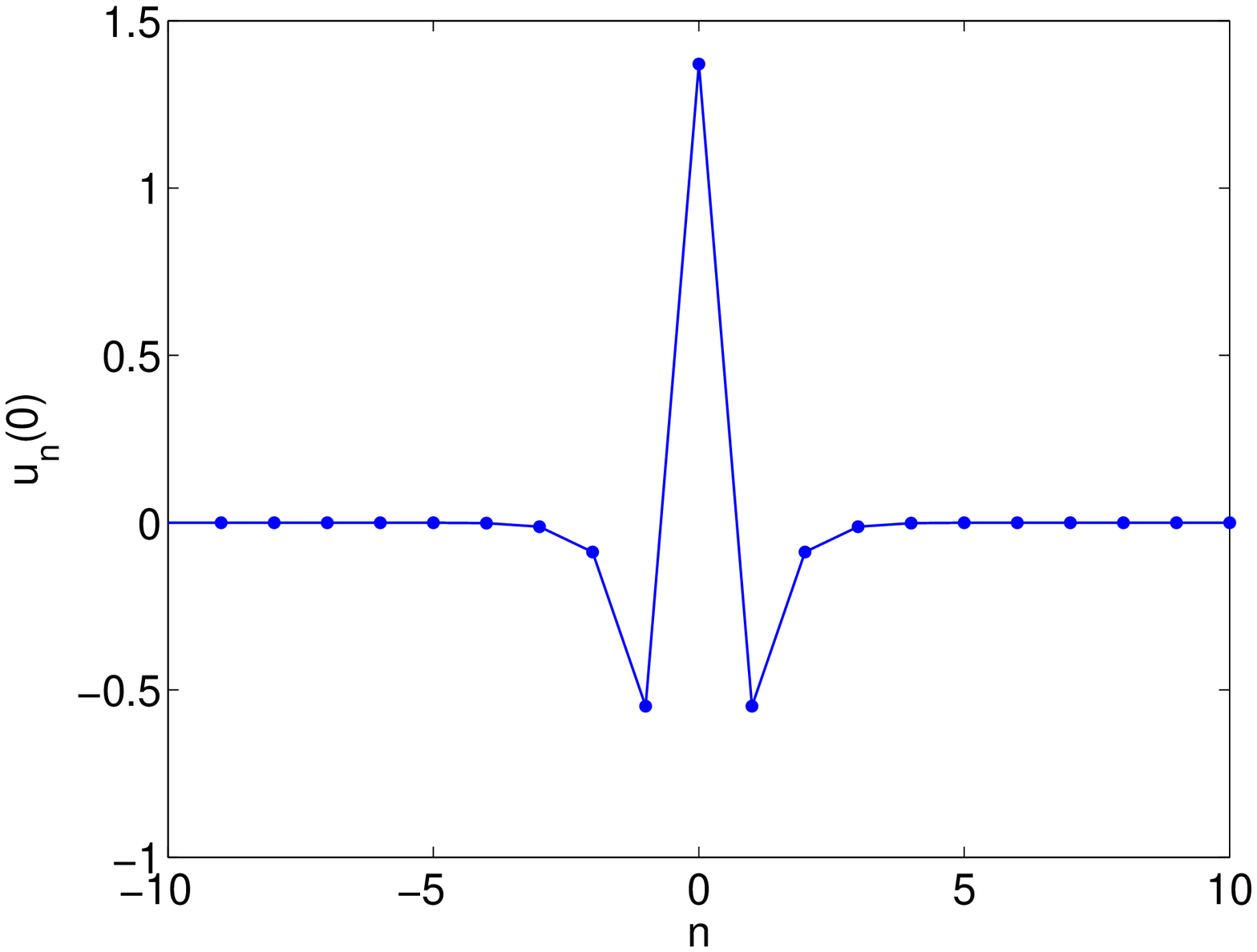} \\
    \includegraphics[width=\middlefig]{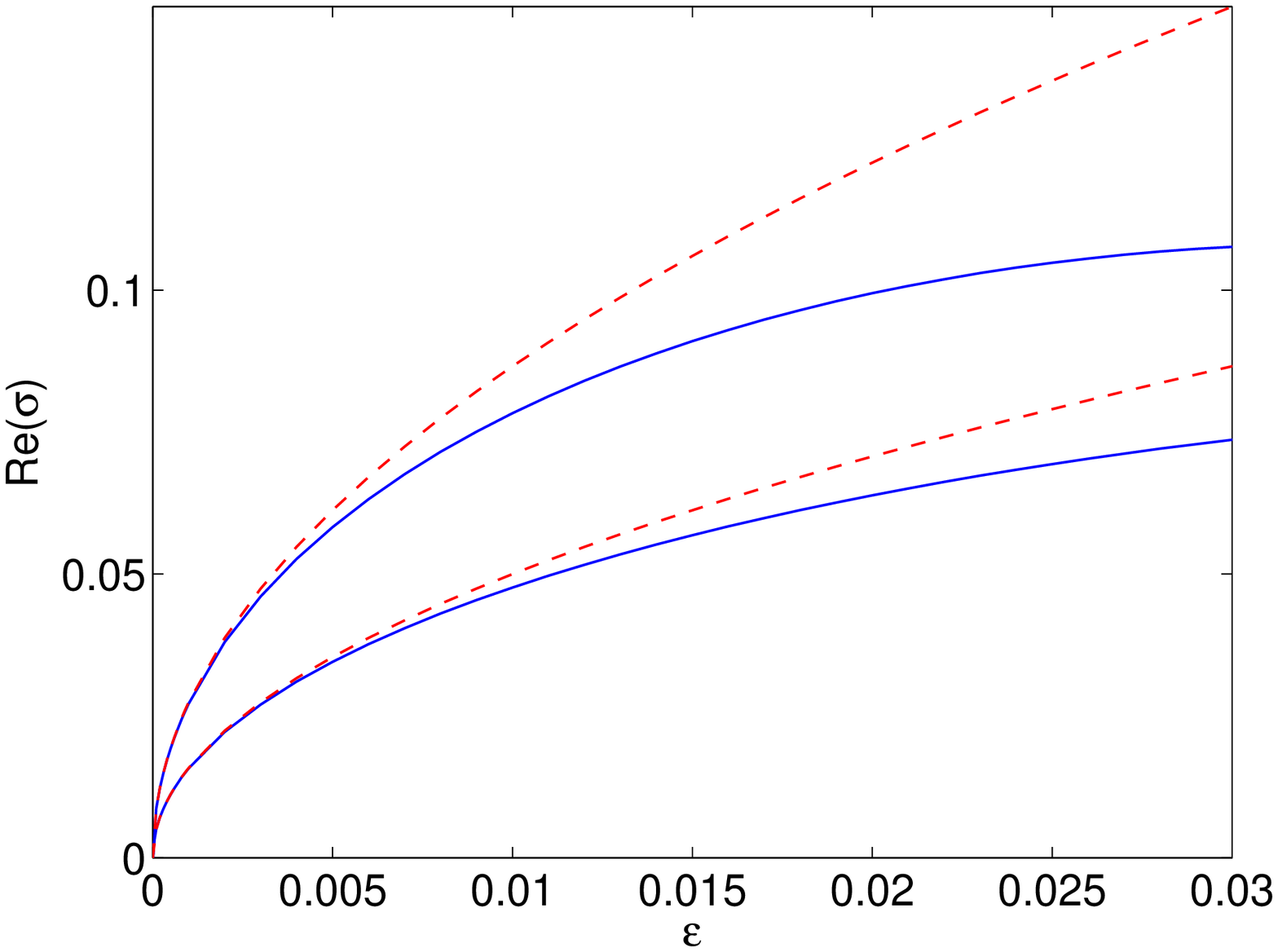} &
    \includegraphics[width=\middlefig]{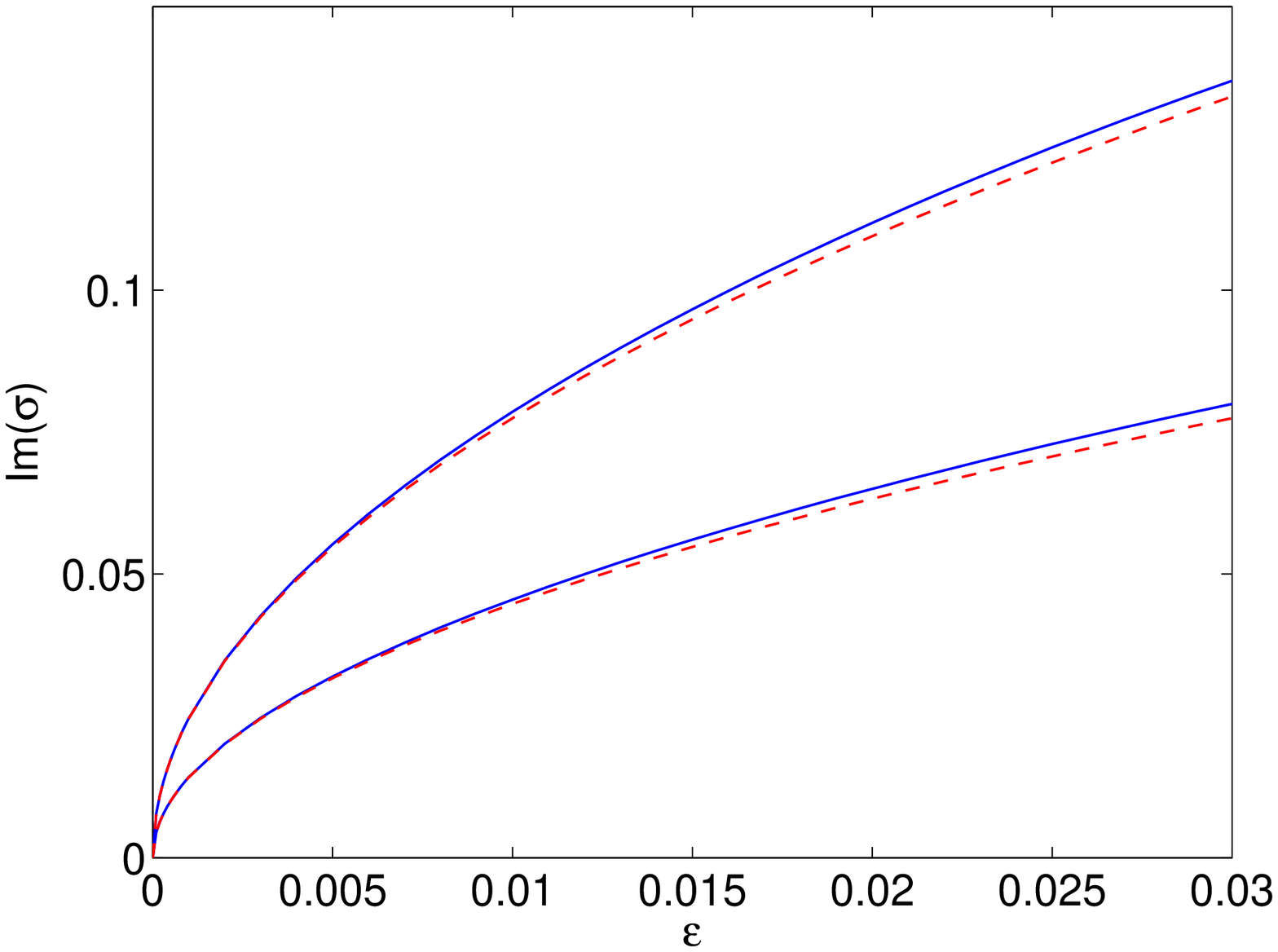} \\
\end{tabular}
\caption{
Same as in Fig. \ref{fig:2site}, but now for the unstable left-panel
configuration of three in-phase excited sites (with two real multiplier
pairs as shown
in the bottom panel) and the potentially stable, close to the anti-continuum
limit, case of the out-of-phase, three-site right-panel configuration.}
\label{fig:3site}
\end{center}
\end{figure}

Figures \ref{fig:2site} and \ref{fig:3site} show, respectively, the
analytical eigenvalue predictions (dashed lines)
for stable and unstable two-site and
three-site breathers with the Morse potential and how they
favorably compare to the corresponding numerical results (solid lines),
obtained via a fully numerical linear stability analysis (and corresponding
computation of the Floquet multipliers). It is clear that the predictions
are very accurate close to the anti-continuum limit, and their validity
becomes progressively limited for larger values of the coupling
parameter $\e$, yet they yield a powerful qualitative and even
quantitative (in the appropriate parametric regime) tool for
tracking the stability of these localized modes.
The figures also illustrate
typical profiles of the corresponding two- and three-site ILMs.

\subsection{Vortices in square lattices}

The methodology can also be extended to lattices of higher dimensionality. We
consider below some basic properties of discrete vortex
breathers of different integer
topological charges, in a square 2D lattice.
Firstly, we consider square vortices over a single
``plaquette'' of the 2D lattice with $S=1$, i.e., at the
anti-continuum limit, the excited sites are
$(0,0)$, $(0,1)$, $(1,1)$ and $(1,0)$ with a phase difference
$\phi=\pi/2$ between nearest neighbors. This implies a perturbation
matrix given by:

\begin{equation}
    Q=q(\pi/2)\tilde{Q},\textrm{with }
    \tilde{Q}=\left(\begin{array}{cccc}
    2 & -1 & 0 & -1 \\
    -1 & 2 & -1 & 0 \\
    0 & -1 & 2 & -1 \\
    -1 & 0 & -1 & 2 \\
    \end{array}\right)
\end{equation}

with $q(\pi/2)$ given from (\ref{eq:qphimorse}) which is evaluated as:

\begin{equation}
    q(\pi/2)=-\frac{(1-\w)^2}{J(1+\w^2)}.
\end{equation}

This, in turn, upon use of Eq. (\ref{floq}) implies that

\begin{equation}
    \sigma=i(1-\w)\sqrt{\frac{\epsilon}{1+\w^2}\tchiQ},
\end{equation}

where $\tchiQ$ are the eigenvalues of the $\tilde{Q}$ matrix. This
corresponds to the matrix of the normal modes of a 1D chain of 4 linearly
coupled oscillators with periodic boundary conditions. Let us recall that for
a system of $n$ coupled oscillators, the eigenvalues are given by:

\begin{equation}
    \tchiQ=4\sin^2\frac{m\pi}{n}\qquad m=1,\ldots,n-1,
\end{equation}
in addition to the $0$ eigenvalue.
In the present case of 4 oscillators with periodic boundary conditions, the nonzero
eigenvalues are given by $2$ and $4$, with the former being doubly degenerate. Thus, we have for $S=1$ vortices the
following spectrum:

\begin{equation}
    \sigma=\left\{\begin{array}{ll}
    i(1-\w)\sqrt{2\frac{\epsilon}{1+\w^2}} & \textrm{single eigenvalue} \\ \\
    2i(1-\w)\sqrt{\frac{\epsilon}{1+\w^2}} & \textrm{double eigenvalue}
    \end{array}\right.
\label{s1p}
\end{equation}
which implies stability for $\epsilon>0$.

This type of analysis can be generalized for arbitrary values of
the vorticity $S$, leading to the conclusion that $\tilde{Q}$ is the matrix
of $4S$ coupled oscillators, which implies that vortices with {\em any}
integer topological charge will be stable for $\epsilon>0$
in the case of a lattice with an on-site Morse potential.
For instance, in the case of the $S=2$ vortex, we obtain the explicit
expressions for the eigenvalues:

\begin{equation}
    \sigma=\left\{\begin{array}{ll}
    \frac{i(1-\w)}{2}\sqrt{(2-2^{1/2})\frac{\epsilon}{1+\w^2}} & \textrm{single eigenvalue} \\ \\
    \frac{i(1-\w)}{2}\sqrt{(2+2^{1/2})\frac{\epsilon}{1+\w^2}} & \textrm{double eigenvalue} \\ \\
    i(1-\w)\sqrt{2\frac{\epsilon}{1+\w^2}} & \textrm{double eigenvalue} \\ \\
    2i(1-\w)\sqrt{\frac{\epsilon}{1+\w^2}} & \textrm{single eigenvalue}
    \end{array}\right.
\label{s2p}
\end{equation}

It is important to highlight here some interesting differences
between the above results and
the case of the DNLS (and more generally that of even
potentials in KG chains, including the case of the hard $\phi^4$ lattice
considered below). In the latter class of problems, the vanishing of the
odd coefficients in the Fourier expansion of the periodic
orbit leads to the conclusion that $q(\pi/2)=0$ and hence
there is no contribution to the eigenvalues to leading
order. This is the situation which has been characterized
as ``super-symmetric'' in \cite{pgkbook} and one in which
the higher order contributions would be critical in determining
the stability. Nevertheless, in the case considered herein,
the asymmetry of the Morse potential produces a nonvanishing
of $q(\pi/2)$ and offers a corresponding nonzero leading order
correction to the eigenvalues at O$(\e^{1/2})$.

\begin{figure}
\begin{center}
\begin{tabular}{cc}
    \includegraphics[width=\middlefig]{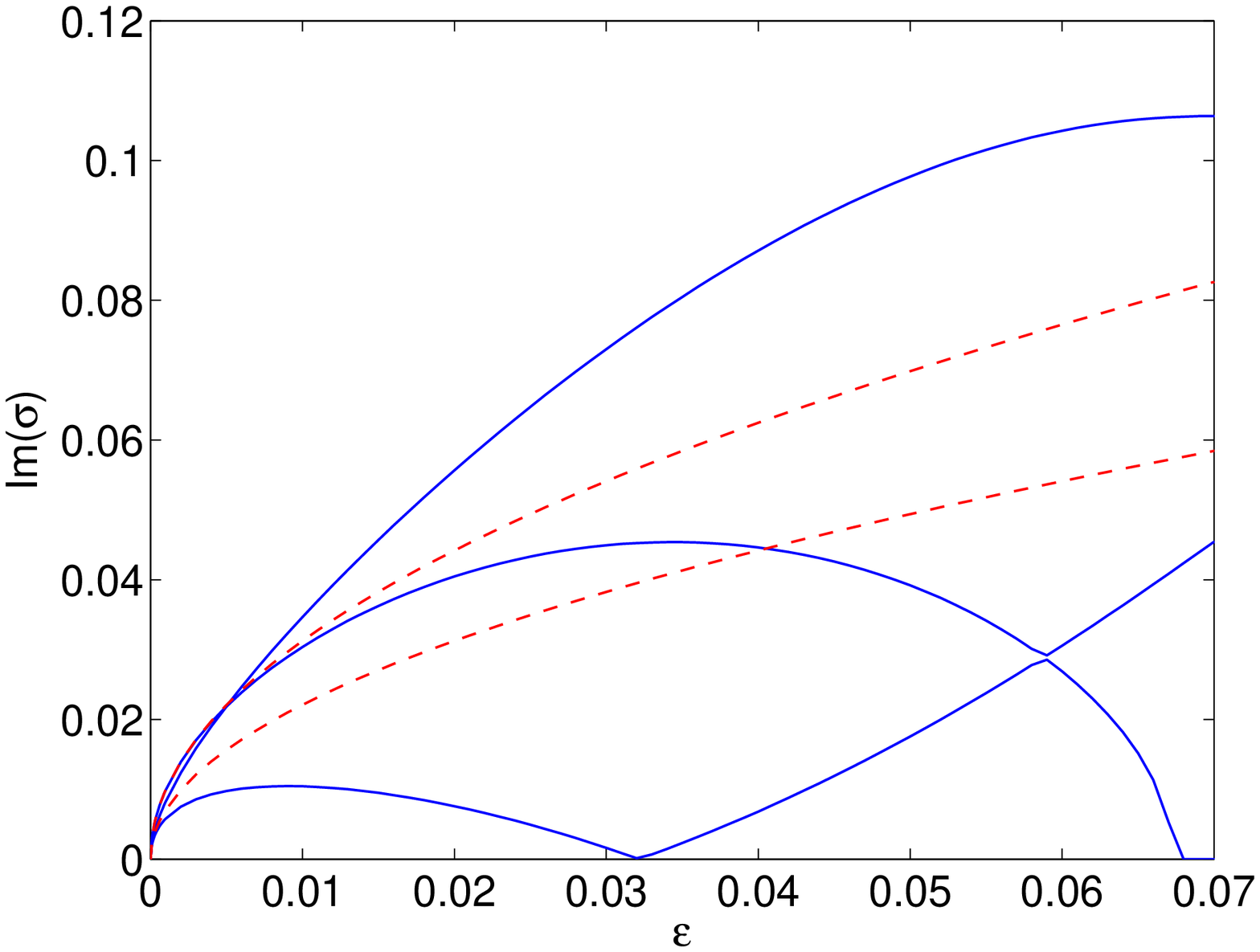} &
    \includegraphics[width=\middlefig]{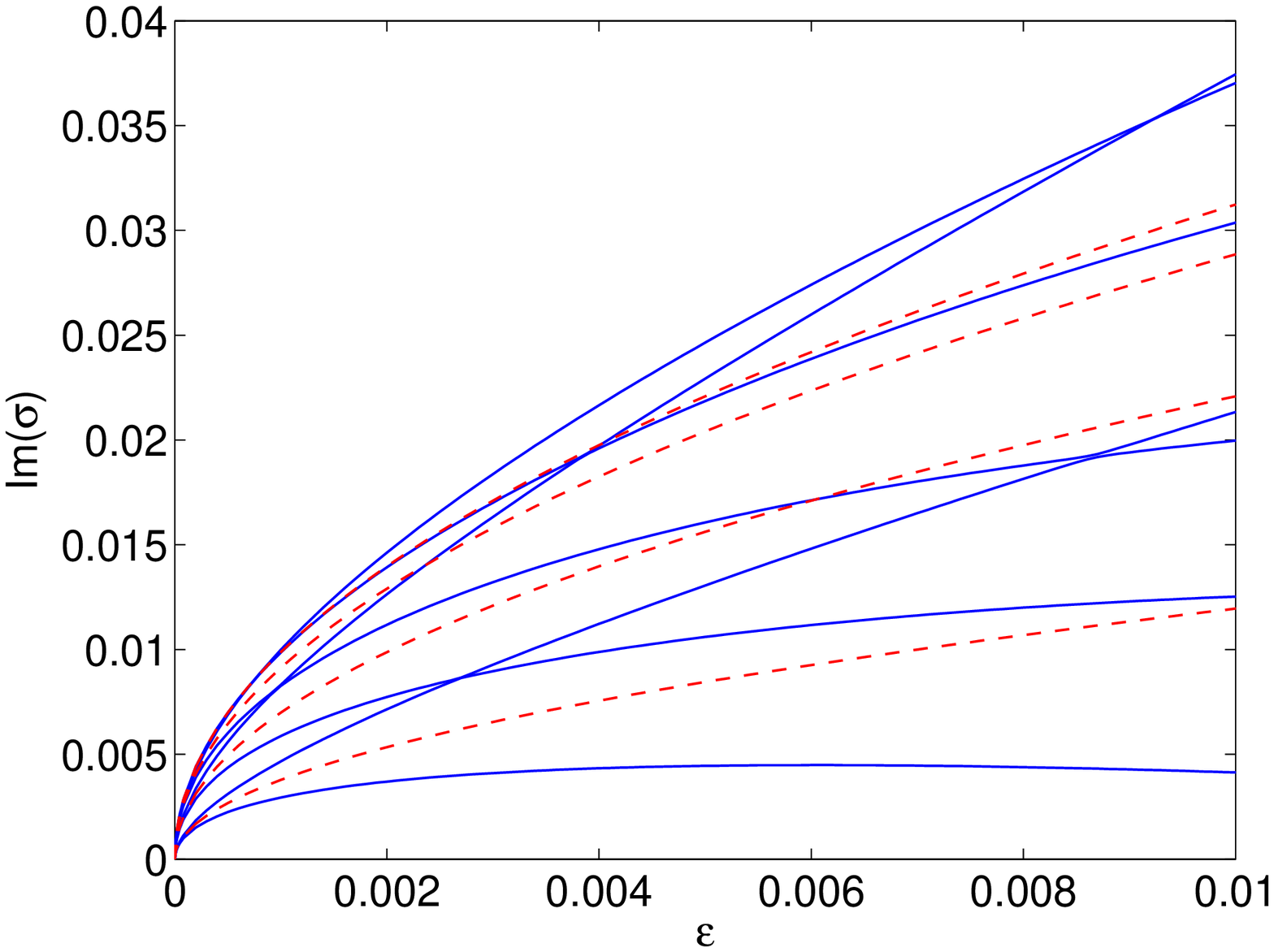} \\
\end{tabular}
\caption{The characteristic exponents of vortex configurations with
$S=1$ (left) and $S=2$ (right), with respect to $\e$, for the
Morse potential
and $\w=0.8$. Dashed lines correspond to the theoretical
predictions based on Eqs. (\ref{s1p}) and (\ref{s2p}), respectively;
the full numerical linear stability results are given by solid lines
and indicate that all doubly degenerate eigenvalue pairs split due
to higher order contributions in the relevant expansions in the
coupling constant $\e$.} \label{fig:vortex}
\end{center}
\end{figure}

Figure \ref{fig:vortex} shows the dependence of stability eigenvalues for
the $S=1$ and $S=2$ vortices and their comparison with the obtained fully
numerical linear stability results as a function of the coupling $\e$. As
can be observed in the figures,
the approximation is less accurate in this case, although it is qualitatively
correct. The reason for the partial disparity is that higher order
contributions to the relevant eigenvalues (whose calculation is considerably
more technically involved) lead to the observed splitting of all the doubly
degenerate eigenvalue pairs. In the relevant cases, the analytical
(dashed line) predictions can be seen to straddle the two observed
numerical pairs.

\section{The Case Example of the Hard $\phi^4$ Potential}

The time evolution of a single oscillator in the hard $\phi^4$ potential,
$V(x)=x^2/2+x^4/4$ is given by:

\begin{equation}
x(t)=\sqrt{\frac{2m}{1-2m}}\cn\left(\frac{ t}{\sqrt{1-2m}},m\right)=
    \sqrt{\frac{2m}{1-2m}}\cn\left(\frac{2K(m)}{\pi}\w t,m\right),
\end{equation}
where $\cn$ is a Jacobi elliptic function of modulus $m$ and
$K(m)$ is the complete elliptic integral of the first kind
defined as
$K(m)=\int_{0}^{\pi/2}\,[1-m\sin^2x]^{-1/2}\ \mathrm{d} x$.

The breather frequency $\w$ is related to the modulus $m$
through:

\begin{equation}\label{eq:trascend4}
    \w=\frac{\pi}{2\sqrt{1-2m}K(m)}.
\end{equation}

The elliptic function can be expanded into a Fourier series leading
to~\cite{Abram}:

\begin{equation}\label{eq:phi4Nome}
    z_{2\nu+1}=\frac{\pi}{K(m)}\sqrt{\frac{2}{1-2m}}\,
    \frac{q^{\nu+1/2}}{1+q^{2\nu+1}}, \qquad \nu=0,1,2,\ldots.
\end{equation}
where $q$ is the elliptic Nome which is defined as

\begin{equation}
    q\equiv q(m)=\exp(-\pi K({1-m})/K(m)).
\end{equation}

In order to get $\chiQ(\phi)$ and $\partial\omega/\partial J$, we cannot use (\ref{eq:fi}) and (\ref{eq:qi}) because it is not possible to find a closed
form expression. Instead, we use the integral expression:

\begin{equation}
    f(\phi_i)=\frac{1}{2\pi\w}\int_{0}^{T}\dot{x}_i(t)\dot{x}_{i+1}(t)\,\mathrm{d}t.
\end{equation}

After some manipulations (where it is crucial to apply \cite[identity 171]{Elliptic}), we obtain:

\begin{eqnarray}
    f(\phi) &=&\frac{8K(m)}{\pi^3\w(1-2m)}\left[\cs(a,m)\ns(a,m)[2E(m)-K(m)(1+\dn^2(a,m))] \right]
\nonumber
\\
&-& \frac{8K(m)}{\pi^3\w(1-2m)}\left[{\mathrm{ds}}(a,m)(\cs^2(a,m)+\ns^2(a,m))\mathrm{Z}(a,m)\right]
\label{evaluation}
\end{eqnarray}

where $E(m)$ is the complete elliptic integral of the second kind
defined as $E(m)=\int_{0}^{\pi/2}\,[1-m\sin^2x]^{1/2}\ \mathrm{d} x$,
$\mathrm{Z}(a,m)$ is the Jacobi zeta function and
$a=2K(m)\phi /\pi$.

For the action $J$, a similar manipulation leads to

\begin{equation}
    J=\frac{16K(m)}{3\pi^2}\left[\frac{1-m}{1-2m}K(m)-E(m)\right].
\end{equation}

The derivative of this expression is cumbersome to handle. So, in what follows,
we will work instead with numerically obtained values of $J$ and
$\partial\omega/\partial J$ which are relevant for time-reversible
multibreathers and vortex breathers, as for these cases we need $f(0)$,
$f(\pi)$ and $f(\pi/2)$. As indicated previously, for every even potential,
$z_{2\nu+1}=0$, and, consequently, $f(0)=J/\omega$, $f(\pi)=-f(0)$ and
$f(\pi/2)=0$. This leads to:

\begin{equation}
    \sigma(0)=\sqrt{-\epsilon\frac{J}{\w}\frac{\partial \omega}{\partial J}\chiQ(0)},
    \label{s0_fi4_ex}
\end{equation}

\begin{equation}
    \sigma(\pi)=\sqrt{\epsilon\frac{J}{\w}\frac{\partial \omega}{\partial J}\chiQ(0)}.
    \label{spi_fi4_ex}
\end{equation}

\begin{figure}
\begin{center}
\begin{tabular}{cc}
    \includegraphics[width=\middlefig]{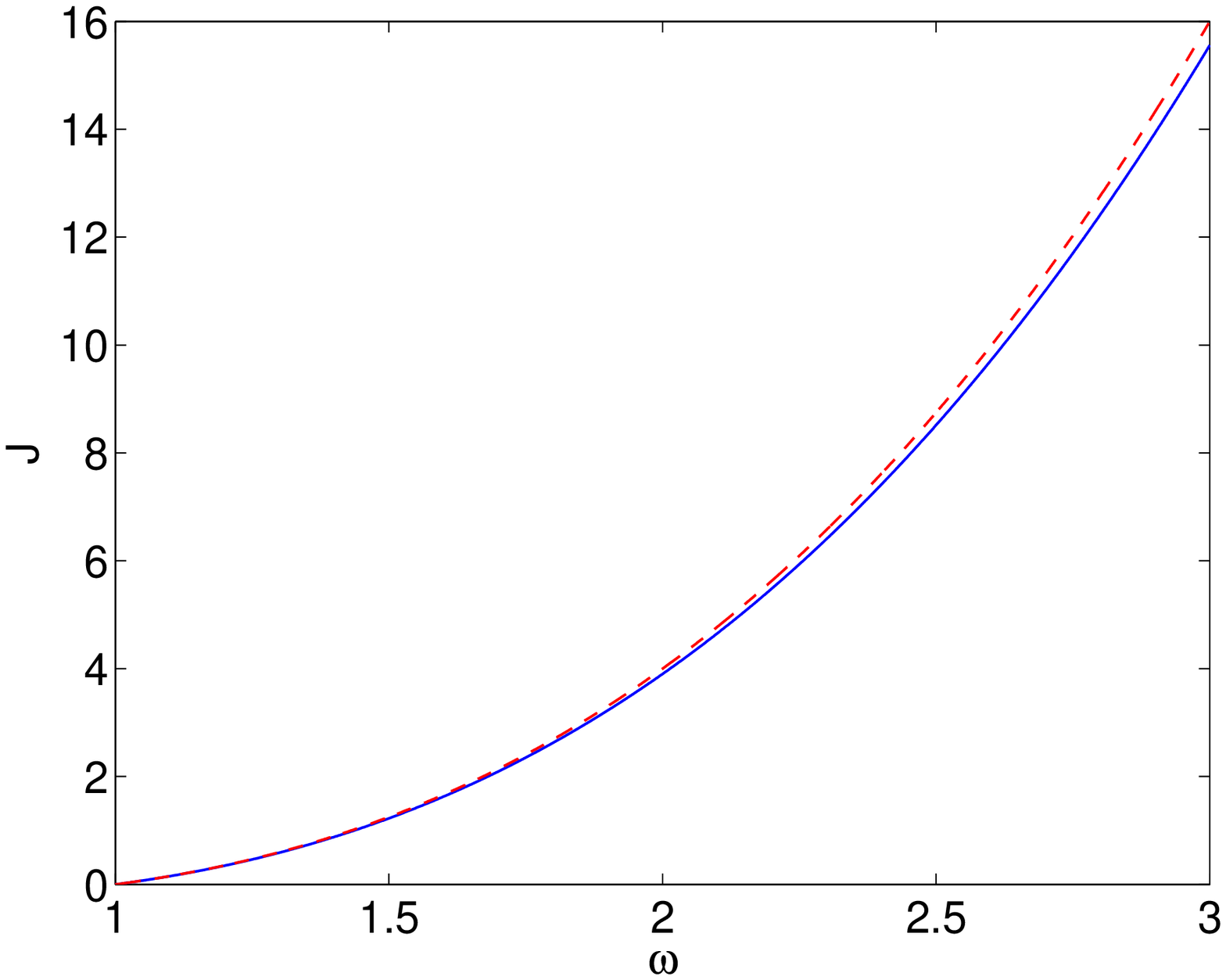} &
    \includegraphics[width=\middlefig]{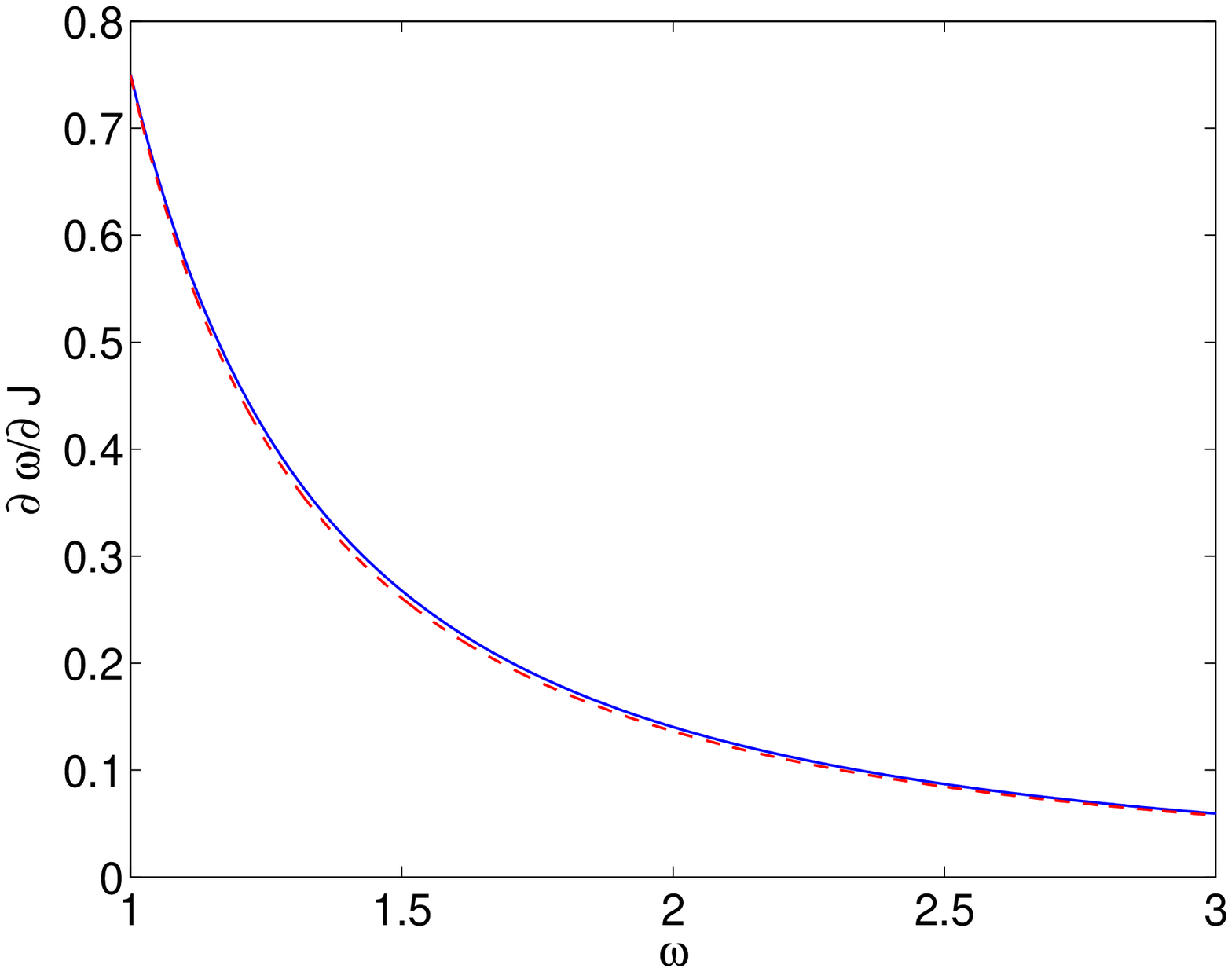} \\
\end{tabular}
\caption{Dependence with respect to $\w$ of the action (left) and
$\partial\w/\partial J$ (right) for the hard $\phi^4$ potential.
The dashed line corresponds to the prediction of the RWA [Eq. (\ref{eq:RWA})],
while the solid one represents the exact numerical result.}
\label{fig:phi4action}
\end{center}
\end{figure}

\begin{figure}[t]
\begin{center}
\begin{tabular}{cc}
    \includegraphics[width=\middlefig]{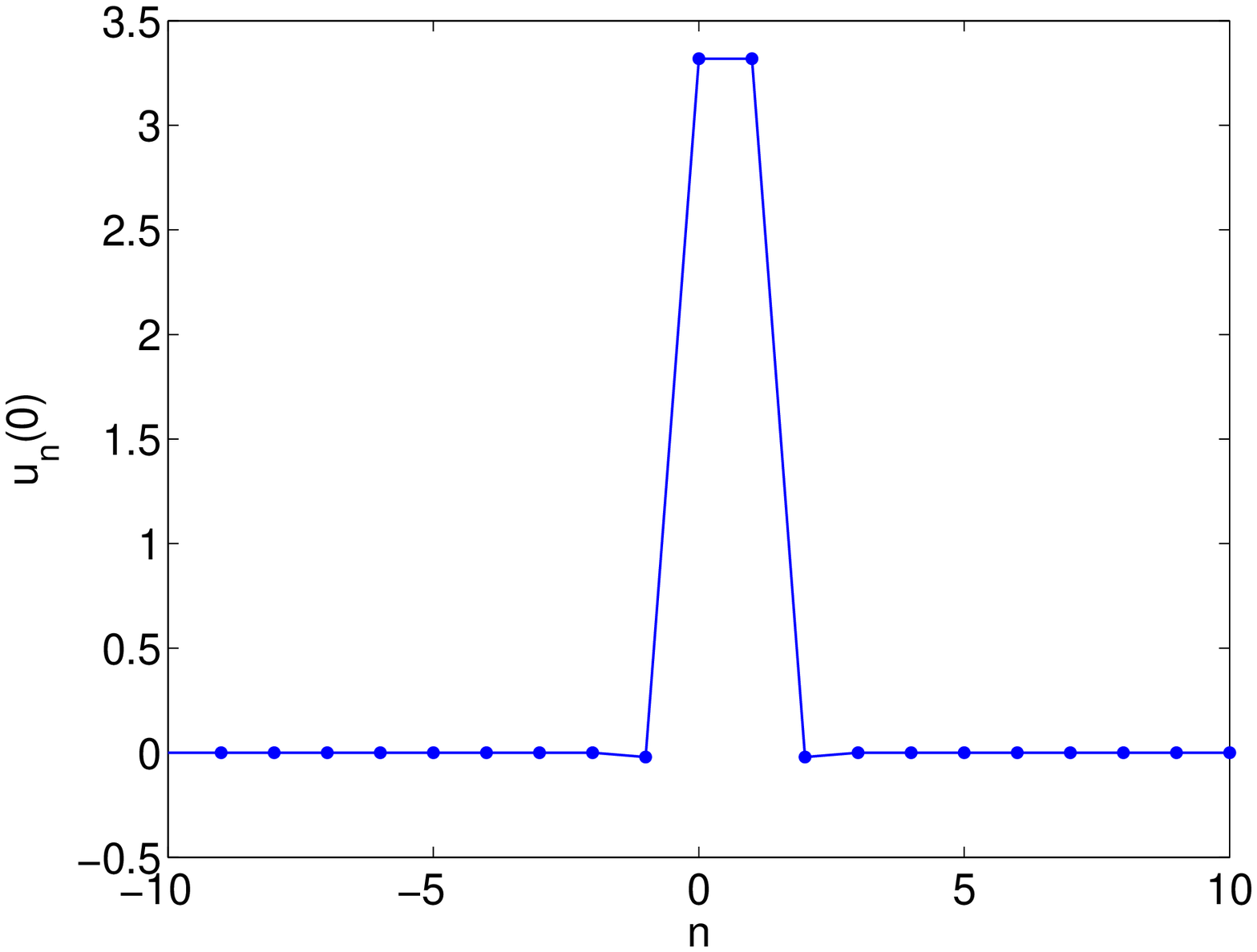} &
    \includegraphics[width=\middlefig]{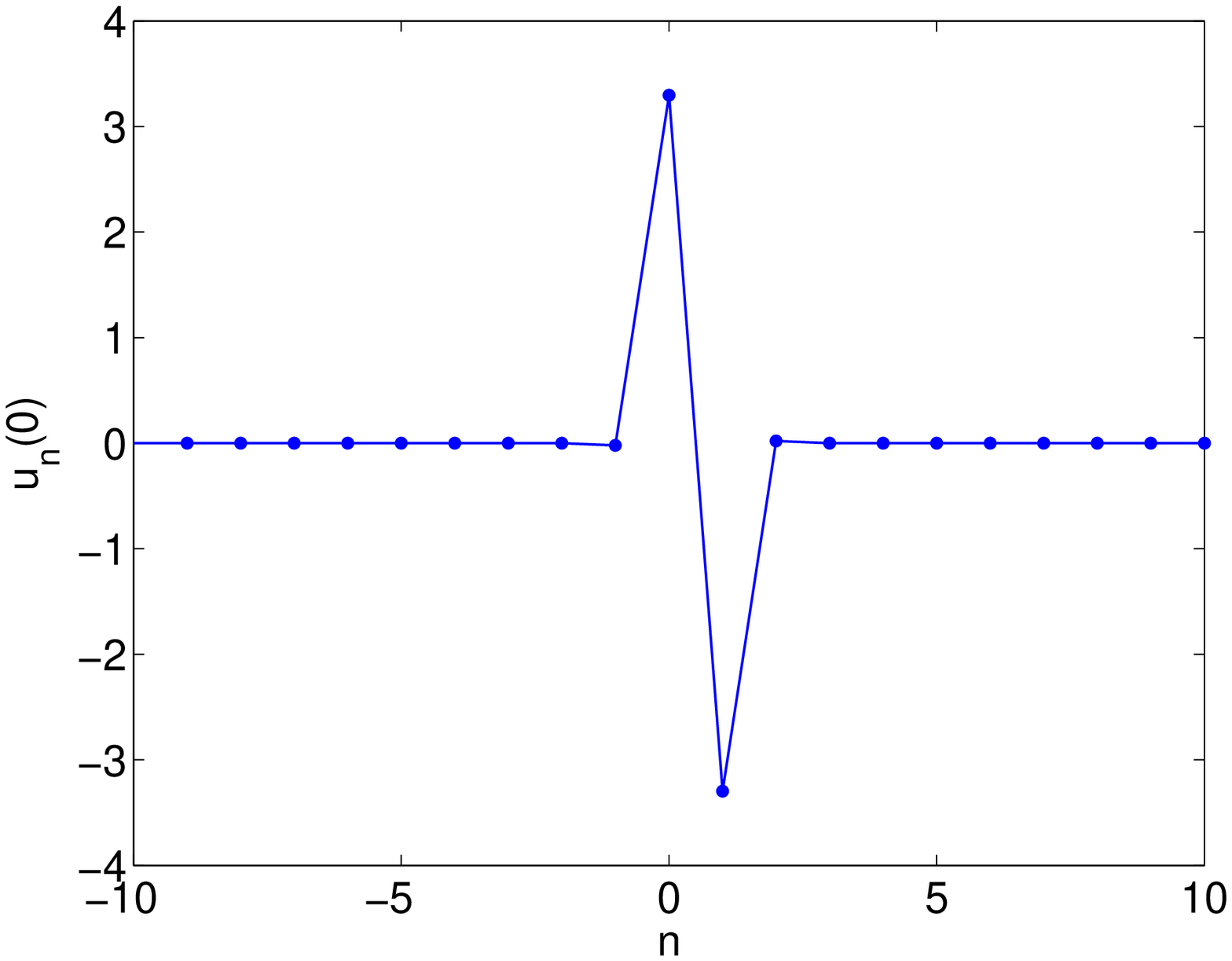} \\
    \includegraphics[width=\middlefig]{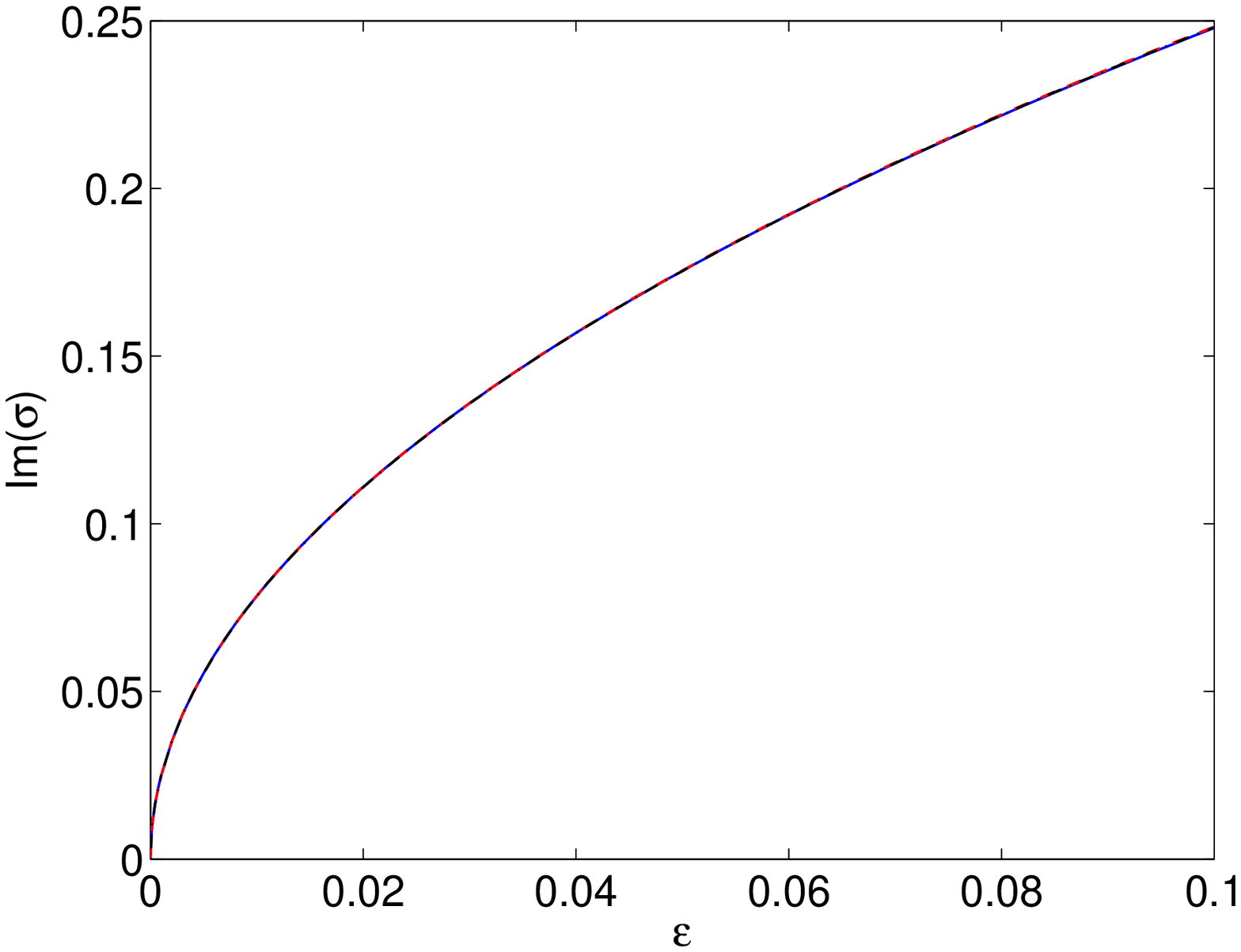} &
    \includegraphics[width=\middlefig]{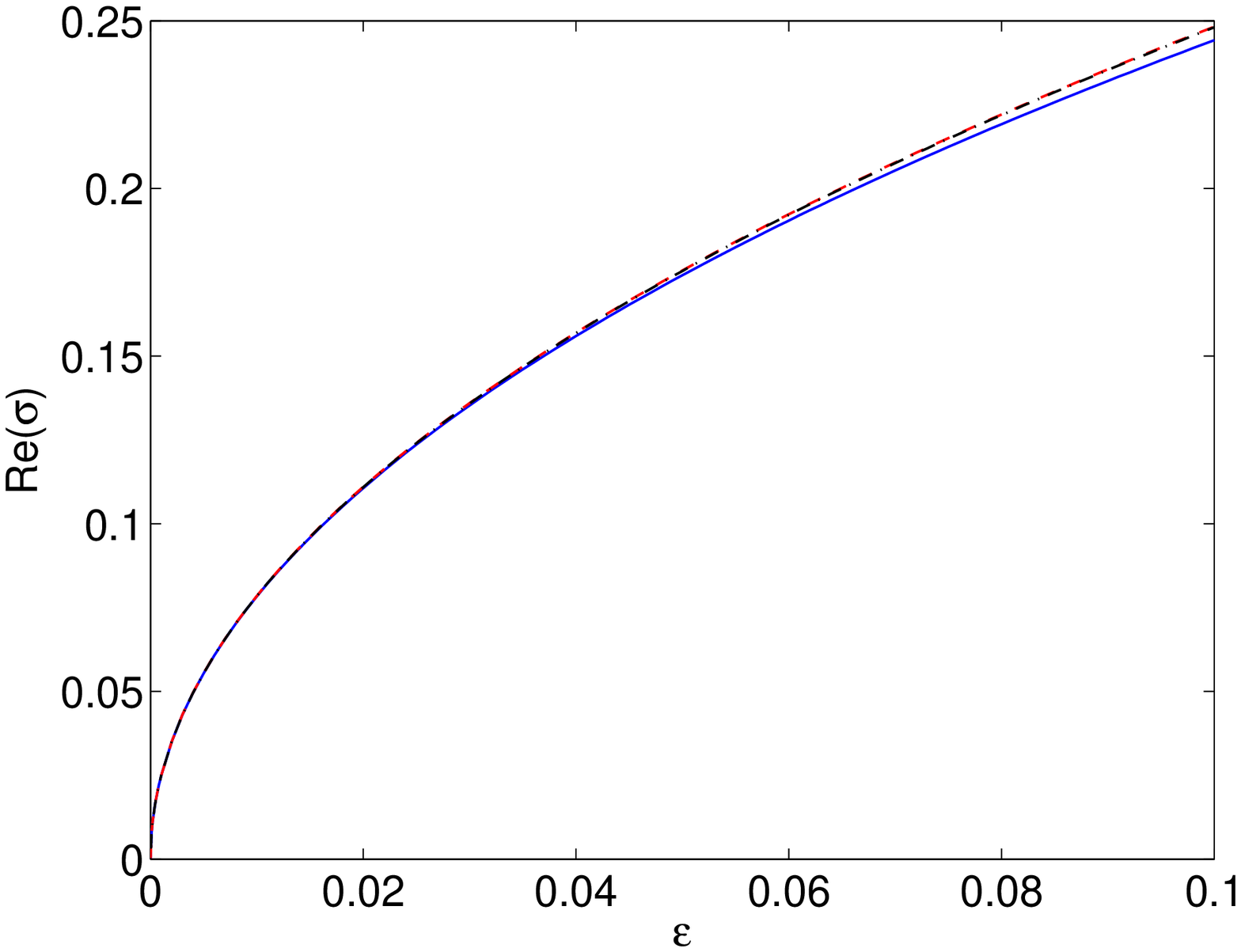} \\
\end{tabular}
\caption{(Top panels) Profiles of an in-phase (left) and an out-of-phase
(right) 2-site breather with the hard $\phi^4$ potential;
$\w=3$ and $\epsilon=0.05$. The bottom panels show the dependence of the
characteristic exponents $\sigma$, of the corresponding configurations,
on the coupling parameter $\e$. The dashed lines correspond to the
predictions of the stability theorems and dash-dotted lines to the RWA
predictions, while the solid ones represent the full numerical
result.} \label{fig:2sitephi4}
\end{center}
\end{figure}

\begin{figure}[t]
\begin{center}
\begin{tabular}{cc}
    \includegraphics[width=\middlefig]{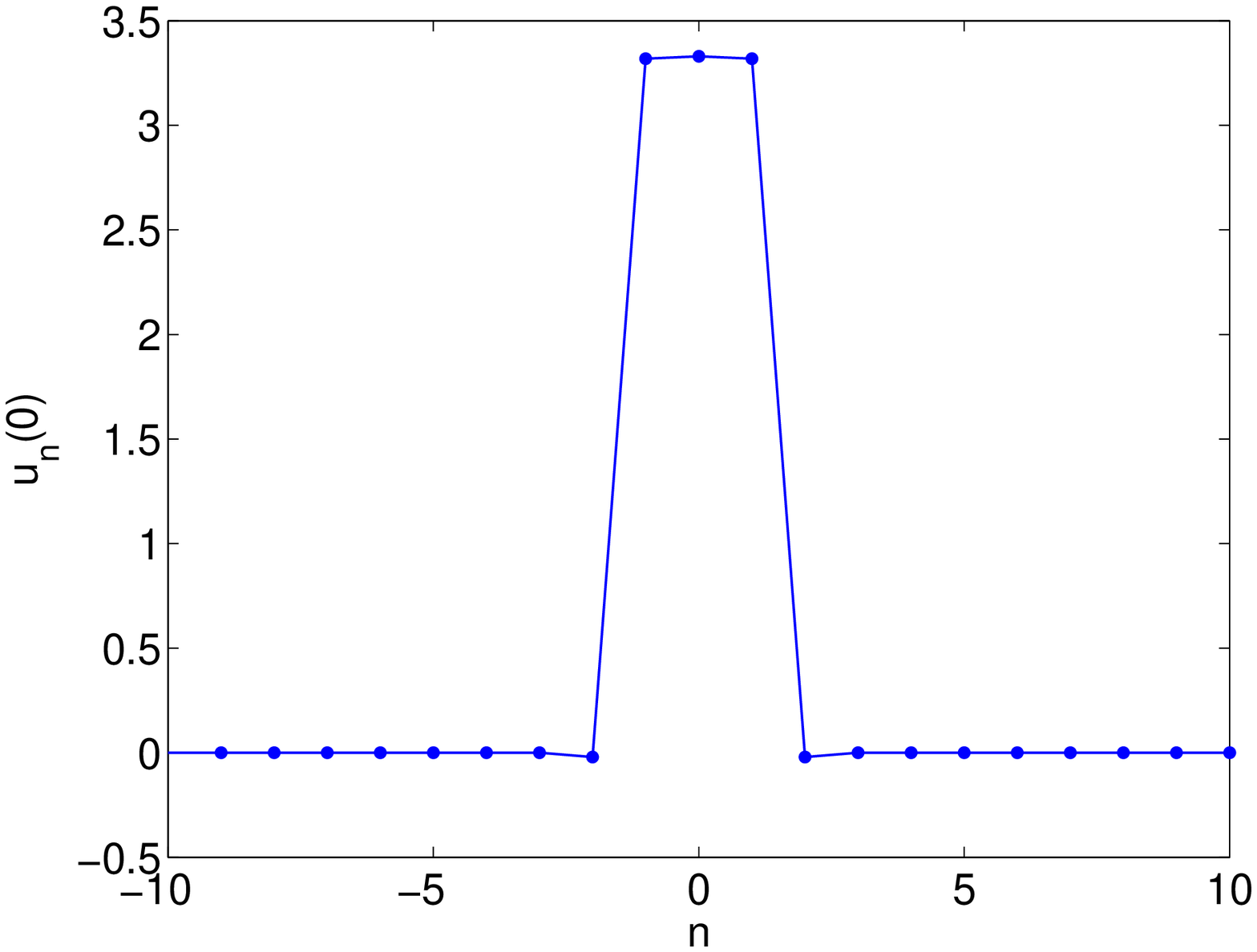} &
    \includegraphics[width=\middlefig]{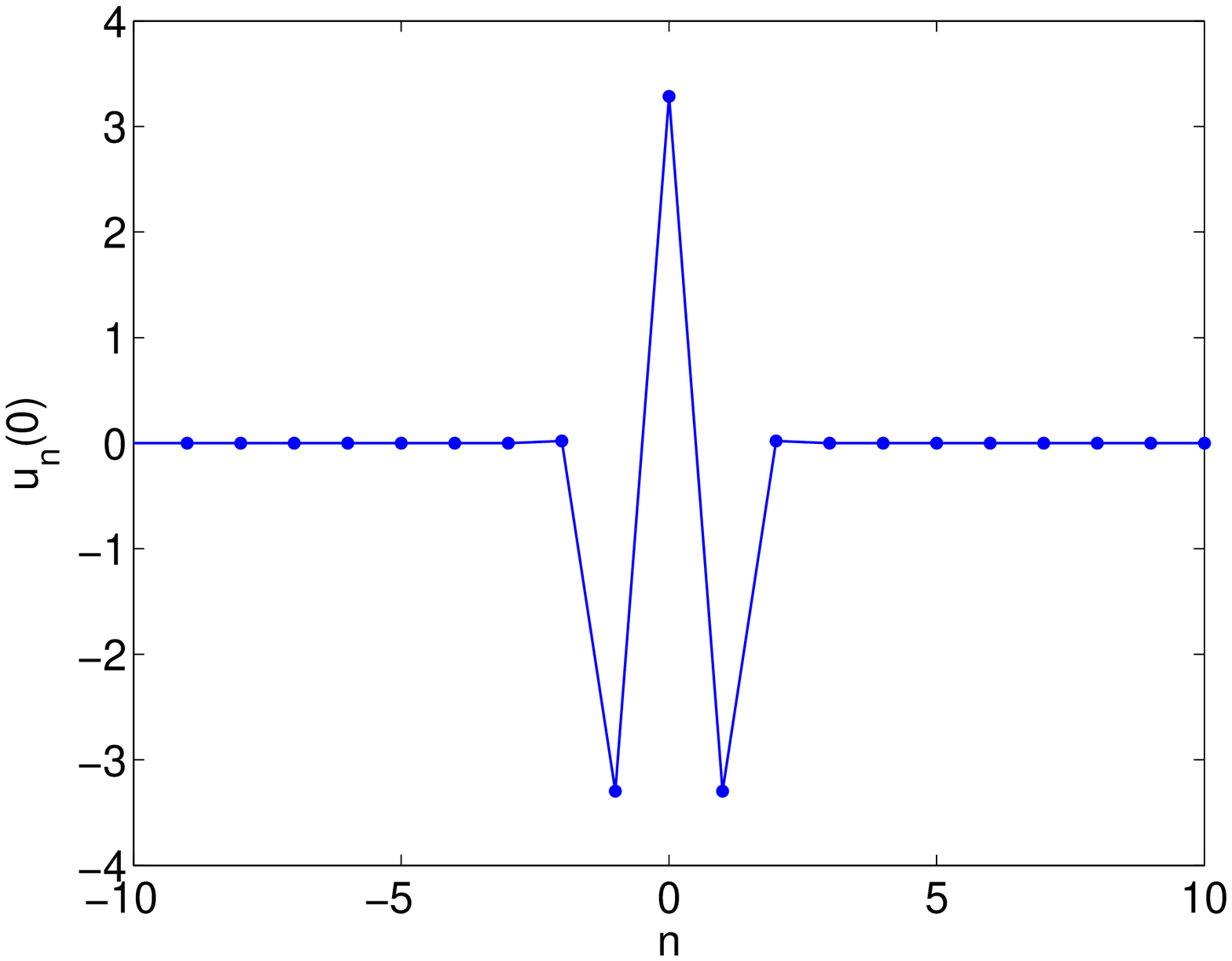} \\
    \includegraphics[width=\middlefig]{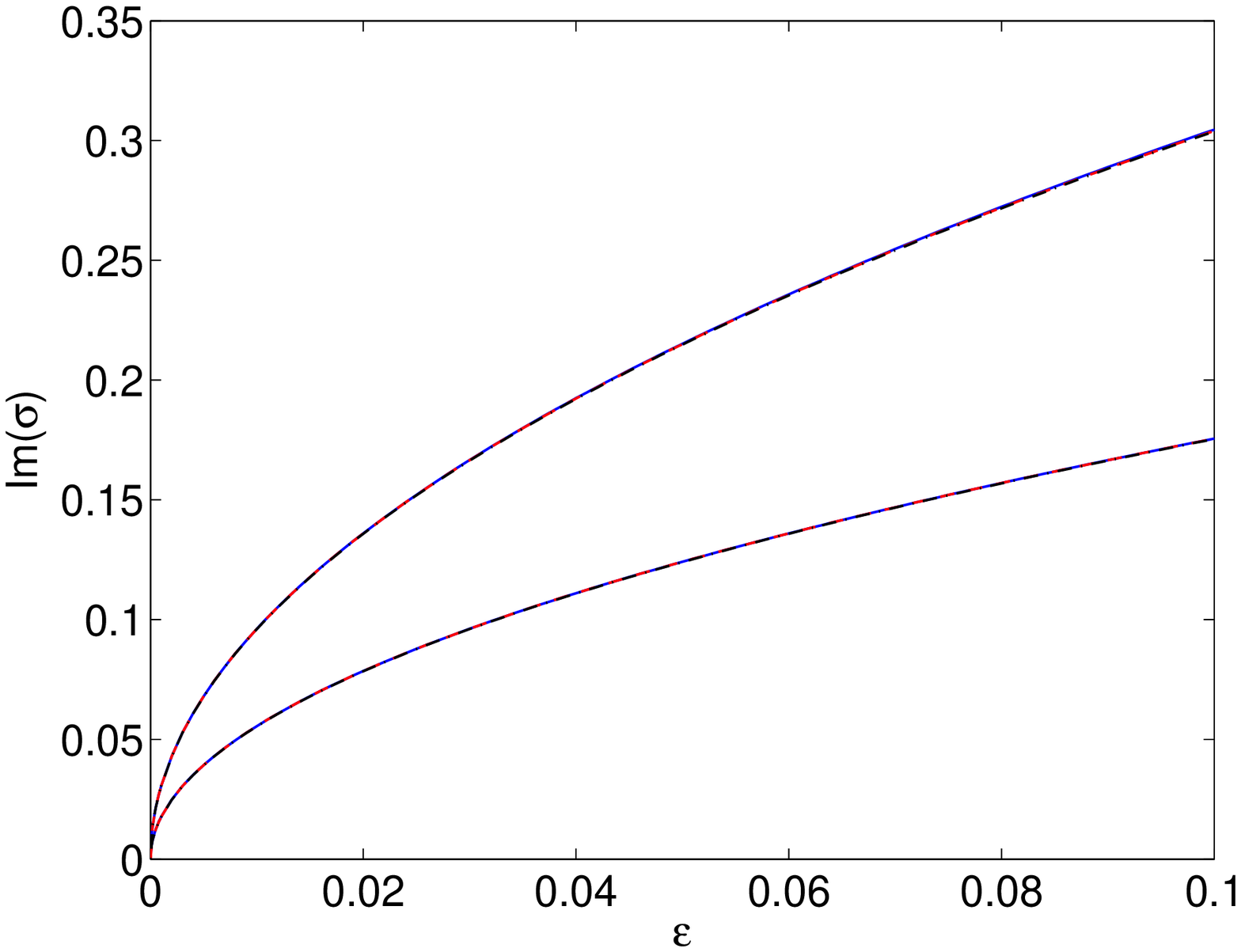} &
    \includegraphics[width=\middlefig]{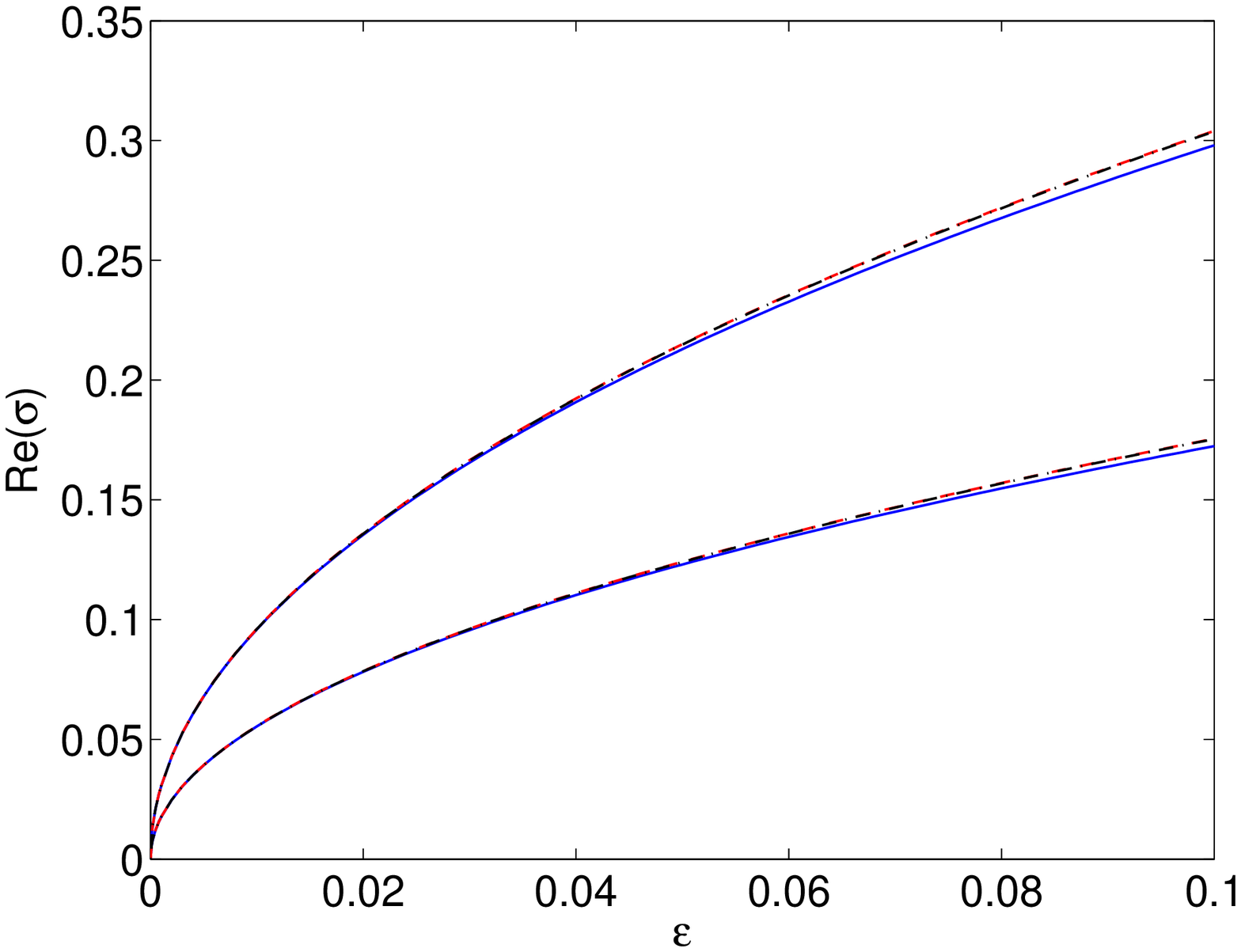} \\
\end{tabular}
\caption{
Same as Fig. \ref{fig:2sitephi4}, but for the three-site
in-phase (left) and out-of-phase (right) configuration.
Again the dash-dotted lines in the bottom panels represent the
(fully-analytical) RWA predictions, which agree well with the semi-analytical
dependences (dashed lines) and, in turn, with the full numerical
results (solid lines).} \label{fig:3sitephi4}
\end{center}
\end{figure}

Figure \ref{fig:phi4action} shows the dependence of $J$ and
$\partial\omega/\partial J$ with respect to the frequency.
Figures \ref{fig:2sitephi4} and \ref{fig:3sitephi4}
illustrate subsequently the relevant stability eigenvalues for 2-site and 3-site breathers as obtained
from the expressions above and compare them to the full numerical
linear stability results. The agreement in this case
is very good (there are no degeneracies and associated higher-order
contributions that may deteriorate the quality of the agreement as
in the vortex breather case above); in fact, in  some of the cases, the curves
are almost indistinguishable throughout the considered parameter
range.

An important observation concerns, however, the role of the ``hard''
nature of the potential. In particular, as illustrated in Fig.
\ref{fig:phi4action}, the quantity $\partial\omega/\partial J$ is
positive in this case, i.e., its sign is opposite from the soft
case of the Morse potential (where $\partial\omega/\partial J=-1$).
This results in the corresponding reversal of the stability conclusions
in Figs. \ref{fig:2sitephi4} and \ref{fig:3sitephi4}, in comparison
with Figs. \ref{fig:2site} and \ref{fig:3site} of the Morse case.
That is, in-phase modes are now stable, while out-of-phase
ones are unstable (as is true for the defocusing nonlinearity
DNLS case also), while the reverse was true in the Morse potential
(as well as for the focusing DNLS case). Lastly, we recall that
since this is an even potential and thus $f(\pi/2)=0$, the leading
order calculation would yield a vanishing contribution to the
eigenvalues for the vortex case and a higher-order calculation
is necessary to determine the stability of the latter.


As an aside towards obtaining a fully analytical prediction for this
case (as some of the quantities need to be obtained numerically above),
we note the following.
Although we cannot acquire an exact form for $J(\w)$, as in the case of the
Morse potential, an approximate form for $J$ can be found by using the
rotating wave approximation (RWA), i.e. by supposing that
$x(t)\approx 2z_1\cos(\w t)$. The introduction in the dynamical
equations for the single oscillator leads to:

\begin{equation}\label{eq:RWA}
    z_1=\sqrt{\frac{\w^2-1}{3}}
\end{equation}

Thus, $J=2\w z_1^2=2\w(\w^2-1)/3$ and $\partial \omega/\partial J=1/[2(\w^2-1/3)]$, and the corresponding expressions for the eigenvalues read:
\begin{equation}
    \sigma(0)\approx\sqrt{-\epsilon\frac{\w^2-1}{3\w^2-1}(\cos\phi)\chiQ(0)}
    \label{s0_fi4_rwa}
\end{equation}

\begin{equation}
    \sigma(\pi)\approx\sqrt{\epsilon\frac{\w^2-1}{3\w^2-1}(\cos\phi)\chiQ(0)}
    \label{spi_fi4_rwa}
\end{equation}

A comparison between the numerically acquired values of $J(\w)$ and $\partial \w/\partial J$ with the ones calculated from the RWA is shown in
figure \ref{fig:phi4action}. The agreement is remarkable
and attests to the quality of the ``single frequency''
rotating wave approximation. In Figs. \ref{fig:2sitephi4} and \ref{fig:3sitephi4} the characteristic
exponents calculated numerically (solid lines)
as well as using Eqs. (\ref{s0_fi4_ex})-(\ref{spi_fi4_ex}) (dashed lines)
and via Eqs. (\ref{s0_fi4_rwa})-(\ref{spi_fi4_rwa}) are compared, illustrating
the excellent agreement between all three.

\section{Conclusions and Perspectives}


The results presented in this work underscore the formulation of
a toolbox that enables the systematic characterization of both
the qualitative and even the quantitative aspects of stability
of multibreather and vortex breather waveforms
in these large number of degree of freedom, Hamiltonian
lattice systems of the Klein-Gordon variety. A systematic
calculation of the corresponding Floquet multipliers is presented
and highlights the crucial components that imply stability,
namely the proper combination of the sign of the coupling
constant, the nature (hard or soft) of the potential and
the relative phases between the adjacent excited sites.
E.g., for positive couplings, and soft potentials, out-of-phase
structures may be stable near the vanishing coupling limit,
while in-phase ones are unstable; the nature of the conclusions
is reversed for either (small) negative couplings or for
hard potentials. The explicit analytical predictions have been
tested against numerical results both for symmetric (such as
the hard $\phi^4$) and asymmetric (such as the Morse) potentials,
both for hard and soft ones, and both for simpler, non-degenerate
one-dimensional multibreather settings and for more complex
and degenerate two-dimensional vortex breathers. In all cases,
the two theories whose results were shown to be equivalent herein,
namely the Aubry band theory and the MacKay Effective Hamiltonian
method yield excellent qualitative and good quantitative agreement
with the full numerical linear stability results. The latter may
not be true only in degenerate cases where higher order contributions
may be critical in breaking the relevant degeneracy (as we saw in
the case of the discrete vortices for the Morse model).

Naturally, a number of interesting directions for future consideration
hereby arise. Perhaps the canonical one among them would involve
a systematic derivation of higher order corrections for prototypical
cases where the leading order approach yields vanishing results.
For instance, the characterization of the stability of discrete
vortices in the ``super-symmetric'' case of phase difference
$\phi=\pi/2$ for even potentials would be a natural example.
Another possibility that is also emerging and would be relevant
to consider from a mathematical point of view would be to examine
models with inter-site nonlinearities, such as ones of the
Fermi-Pasta-Ulam type. In these cases, where the potential
is a function $V(x_{n}-x_{n-1})$, it is relevant to point
out that upon consideration of the so-called strain variables
$r_n=x_n-x_{n-1}$, the problem is reverted to an on-site potential
case, for which it would be worthwhile to explore methods similar
to the ones analyzed herein. These directions are presently
under consideration and will be reported in future publications.

{\it Acknowledgments.} PGK gratefully acknowledges support from
NSF-DMS-0349023 (CAREER), NSF-DMS-0806762 and from the Alexander
von Humboldt Foundation. JC and JFRA acknowledge financial support
from the MICINN project FIS2008-04848.

\appendix
\renewcommand\thesection{Appendix \Alph{section}}

\newcommand{\noopsort}[1]{} \newcommand{\printfirst}[2]{#1}
  \newcommand{\singleletter}[1]{#1} \newcommand{\switchargs}[2]{#2#1}


\end{document}